\documentclass[aps,prl,reprint,twocolumn,showpacs,showkeys,footinbib,amsmath,amssymb,superscriptaddress]{revtex4-2}

\usepackage{amssymb}
\usepackage{amsfonts}
\usepackage{graphicx}
\usepackage{epstopdf}
\usepackage{dcolumn}
\usepackage{bm}
\usepackage{color}
\usepackage{textcomp}
\usepackage[colorlinks=true,linkcolor=blue,citecolor=blue,anchorcolor=blue,urlcolor=blue]{hyperref}
\begin{document}


\title{Altermagnets Enable Gate-Switchable Helical and Chiral Topological Transport with Spin–Valley–Momentum-Locked Dual Protection}



\author{Xianzhang Chen}
\affiliation{Eastern Institute for Advanced Study, Eastern Institute of Technology, Ningbo, Zhejiang 3150200, P. R. China}
\affiliation{International Center for Quantum Design of Functional Materials (ICQD), \\and Hefei National Laboratory, University of Science and Technology of China, Hefei, 230026, China}

\author{Jiayong Zhang}
\affiliation{Eastern Institute for Advanced Study, Eastern Institute of Technology, Ningbo, Zhejiang 3150200, P. R. China}
\affiliation{International Center for Quantum Design of Functional Materials (ICQD), \\and Hefei National Laboratory, University of Science and Technology of China, Hefei, 230026, China}
\affiliation{School of Physical Science and Technology, Suzhou University of Science and Technology, Suzhou 215009, China}

\author{Bowen Hao}
\affiliation{Eastern Institute for Advanced Study, Eastern Institute of Technology, Ningbo, Zhejiang 3150200, P. R. China}

\author{Jiahui Qian}
\affiliation{Eastern Institute for Advanced Study, Eastern Institute of Technology, Ningbo, Zhejiang 3150200, P. R. China}

\author{Ziye Zhu}
\affiliation{Eastern Institute for Advanced Study, Eastern Institute of Technology, Ningbo, Zhejiang 3150200, P. R. China}
\affiliation{International Center for Quantum Design of Functional Materials (ICQD), \\and Hefei National Laboratory, University of Science and Technology of China, Hefei, 230026, China}


\author{Igor \v{Z}uti\'c}
\address{Department of Physics, University at Buffalo, State University of New York, Buffalo, New York 14260, USA}

\author{Zhenyu Zhang}
\affiliation{International Center for Quantum Design of Functional Materials (ICQD), \\and Hefei National Laboratory, University of Science and Technology of China, Hefei, 230026, China}

\author{Tong Zhou}
\email{tzhou@eitech.edu.cn}
\affiliation{Eastern Institute for Advanced Study, Eastern Institute of Technology, Ningbo, Zhejiang 3150200, P. R. China}
\date{Feburary 20, 2026}

\begin{abstract}

We establish a unified, symmetry-driven framework that combines the alternating spin splitting of altermagnets with valley topology to realize and electrically interconvert helical and chiral topological phases within a single material platform. We first demonstrate a magnetic analogue of the quantum spin Hall effect in altermagnets, hosting helical spin–valley–momentum–locked (SVML) edge states characterized by a composite spin-valley Chern number $C_{\mathrm{sv}}=2$. Large-scale quantum transport simulations show these SVML edge states exhibit fully quantized spin conductance robust against nonmagnetic and long-range magnetic disorder, reflecting their dual topological protection, while remaining vulnerable to short-range magnetic disorder. Exploiting that the counterpropagating SVML modes are linked by crystal rotation symmetry, we introduce a gate-tunable sublattice-staggered potential that selectively gaps one valley and converts the helical state into a chiral quantum anomalous Hall phase with $C_{\mathrm{sv}}=1$, robust against all disorder types. Reversing the potential switches the transmitted spin–valley polarization. Our first-principles calculations identify monolayer V$_2$STeO and VO families as realistic platforms supporting both helical and chiral topological phases and their electrical switching. These results establish altermagnets as electrically programmable platforms for robust topological devices across charge, spin, and valley.

\end{abstract}

\maketitle

The quantum spin Hall (QSH) effect hosts dissipationless helical edge states protected by time-reversal symmetry (TRS)~\cite{Hasan2010:RMP,Qi2011:RMP}. While spin–orbit coupling (SOC) is essential to open the topological gap, it simultaneously entangles spin channels and precludes strictly quantized spin Hall conductivity ~\cite{Hasan2010:RMP,Qi2011:RMP,Matusalem2019:PRB,Monaco2020:PRB}. Moreover, even weak magnetic disorder breaks TRS, destabilizing helical edges and destroying quantized transport~\cite{Maciejko2009:PRL,Kimme2016:PRB,Jack2020:PNAS,Zhou2021:PRL}. A common route to overcome this fragility is to convert QSH phases into quantum anomalous Hall (QAH) phases by introducing magnetism to break TRS~\cite{Yu2010:Science,Chang2013:Science,Chang2023:RMP}. However, this strategy typically requires long-range ferromagnetic order and finely tuned Chern gap reopening, severely constraining material realizations, operational temperatures, and device tunability~\cite{Tokura2019:NRP,Chang2023:RMP}.

\begin{figure}[ht!]
	\includegraphics[width=0.9\linewidth]{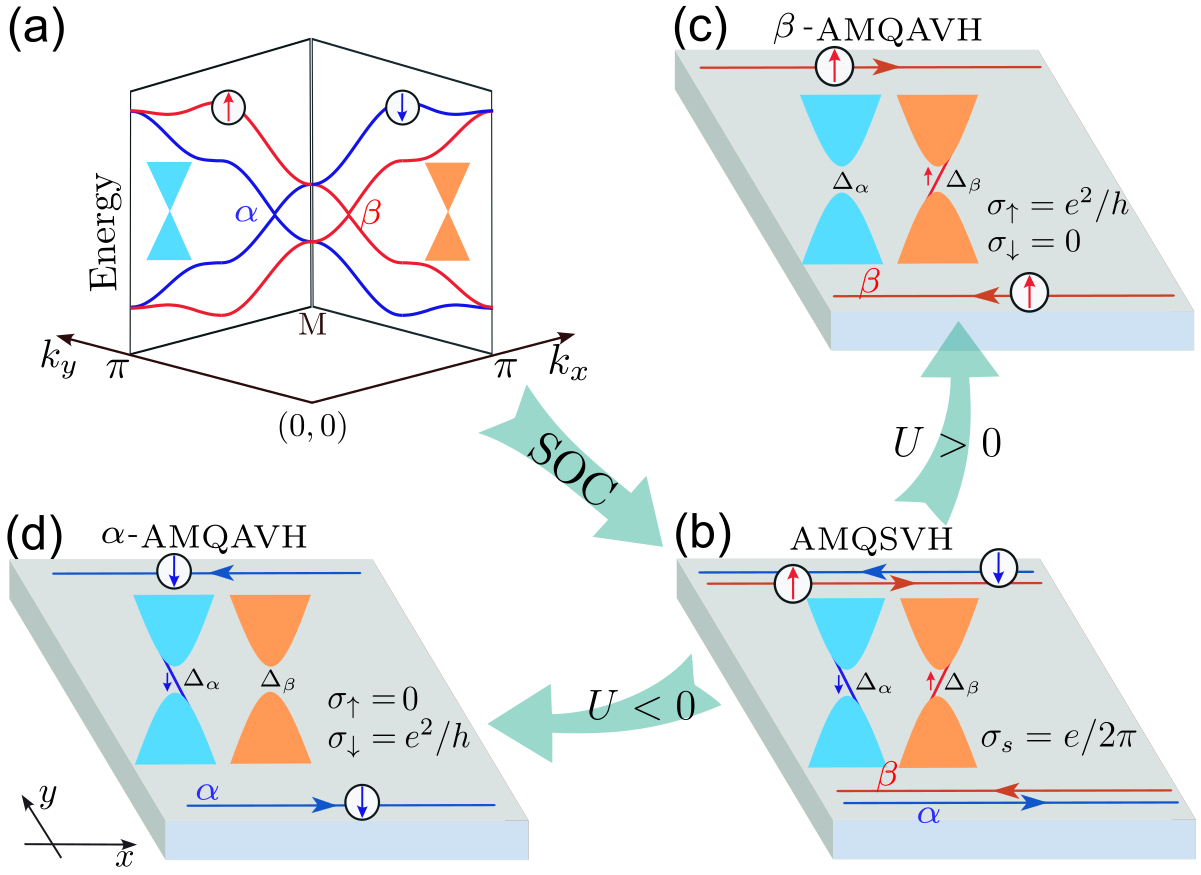}
	\caption{(a) Schematic of spin–valley–resolved Weyl points at the $\alpha$ and $\beta$ valleys in AMs. (b) SOC opens topological gaps at both valleys, yielding an AMQSVH phase with helical SVML edge states and fully quantized spin Hall conductivity ($\sigma_s=e/2\pi$). (c) A positive staggered potential ($U>0$) selectively trivializes the gap at the $\alpha$ valley, eliminates its edge state, and produces a $\beta$-AMQAVH phase with spin-up chiral edge states at the $\beta$ valley ($\sigma_{\uparrow}=e^2/h$). (d) Reversing the staggered potential ($U<0$) yields an $\alpha$-AMQAVH phase with spin-down chiral edge states at the $\alpha$ valley ($\sigma_{\downarrow}=e^2/h$).}
	\label{fig:schmetic_setup}
\end{figure}

The emergence of altermagnets (AMs) offers a promising route to overcome these limitations~\cite{Wu2007:PRB,Hayami2019:JPSJ,Yuan2020:PRB,Smejkal2020:SA,Mazin2021:PNAS, Ma2021:NC, Krempasky2024:Nature,Zhou2025:Nature,Zhang2025:NP,Jiang2025:NP,Wang2025:arXiv,FuD2025:arXiv,Smejkal2022:PRX,Smejkal2022:PRX2}. AMs are collinear antiferromagnets with zero net magnetization yet spin-polarized bands, combining key advantages of ferromagnets and antiferromagnets~\cite{Smejkal2022:PRX, Smejkal2022:PRX2,Bai2024:AFM,Song2025:NRM, Bhowal2025:arXiv,Cheong2025:NQM,Guo2025:AM,Yu2025:AM,Fukaya2025:JPCM,Duan2025:PRL, Zhu2025:NL,Gu2025:PRL, Sun2025:AM,Liu2025:NP,Jungwirth2026:Nature}. Their momentum-space alternating spin polarization naturally separates spin channels by valley and suppresses interchannel spin mixing, offering a rich landscape of robust spin-valley-related altermagnetic topological phases~\cite{Mazin2023:arXiv,Guo2023:NCM,Chen2023:APL,Ma2024:PRB,Tan2025:PRB,Feng2025:arXiv,Antonenko2025:PRL,Zhang2025:arXiv,Shi2026:APL,Fu2025:arXiv,Chen2025:arXiv,Yan2026:PRB,Yang2025:NL,Xu2025:NSR,Wan2025:PRB,Seyler2026:arXiv}. Unlike time-reversal-symmetric topological insulators, AMs intrinsically break TRS, opening new possibilities for realizing anomalous topological transport without invoking additional ferromagnetism~\cite{Bai2024:AFM,Liu2025:NP,Jungwirth2026:Nature}. Moreover, the relevant spin–valley sectors are connected by crystal rotation symmetries (CRS)~\cite{Ma2021:NC,Smejkal2022:PRX}, suggesting a symmetry-driven geometric handle to control topological phases and their transport behaviors. Despite these appealing features, a unified framework to describe and electrically manipulate distinct topological phases and their transport behaviors within a single AM platform remains lacking, particularly for rigorous topological characterization, robustness against disorder, and realistic material/device implementations.

Motivated by these opportunities, we develop a general framework that combines altermagnetic splitting with valley topology to realize and electrically switch helical QSH and chiral QAH states via selective valley gapping (Fig.~\ref{fig:schmetic_setup}). Starting from well-established AM Weyl semimetals ~\cite{Jungwirth2026:Nature,Xu2025:NSR} with two valleys ($\alpha$ and $\beta$) [Fig.~\ref{fig:schmetic_setup}(a)], we show that SOC opens topological gaps at both valleys, yielding an altermagnetic QSH (AMQSVH) phase with spin–valley–momentum-locked (SVML) edge states as in Fig.~\ref{fig:schmetic_setup}(b). Unlike conventional QSH systems with only spin–momentum locking, SVML provides dual protection captured by a composite spin–valley Chern number  $C_{\mathrm{sv}}=2$. Large-scale transport simulations confirm the AMQSVH phase keeps robust quantized conductance under nonmagnetic disorder and long-range magnetic fluctuations, reflecting its dual topological protection, but it is degraded by short-range (e.g. Anderson-type) magnetic disorder that can couple counterpropagating SVML channels.

Exploiting that the helical AMQSVH edge states are related by CRS, breaking this symmetry selectively gaps out one partner of the helical pair, thereby converting the AMQSVH into a chiral QAH phase without introducing ferromagnetism. We realize this transformation through a sublattice-staggered potential, $U$, which is natural in two-sublattice 2D materials and can be continuously controlled by an electric field~\cite{schaibley2016:NRM,Li2016:NN,Li2018:Science,Ren2016:RPP}. This gate-tunable $U$ enables selective valley gapping: by adjusting its strength, one valley is rendered topologically trivial while the other remains nontrivial, yielding a chiral SVML altermagnetic QAH (AMQAVH) phase with $C_{\mathrm{sv}}=1$, featuring gate-selectable spin–valley polarization, and robustness against all disorder types [Figs.~\ref{fig:schmetic_setup}(c)-(d)]. Guided by this symmetry–topology framework, our first-principles calculations further identify realistic monolayer families, including V$_2$STeO and VO that support electrical switching between AMQSVH and AMQAVH phases. These results establish AM as a versatile platform for topological transport across charge, spin, and valley degrees of freedom.

\begin{figure}[t]
	\includegraphics[width=0.9\linewidth]{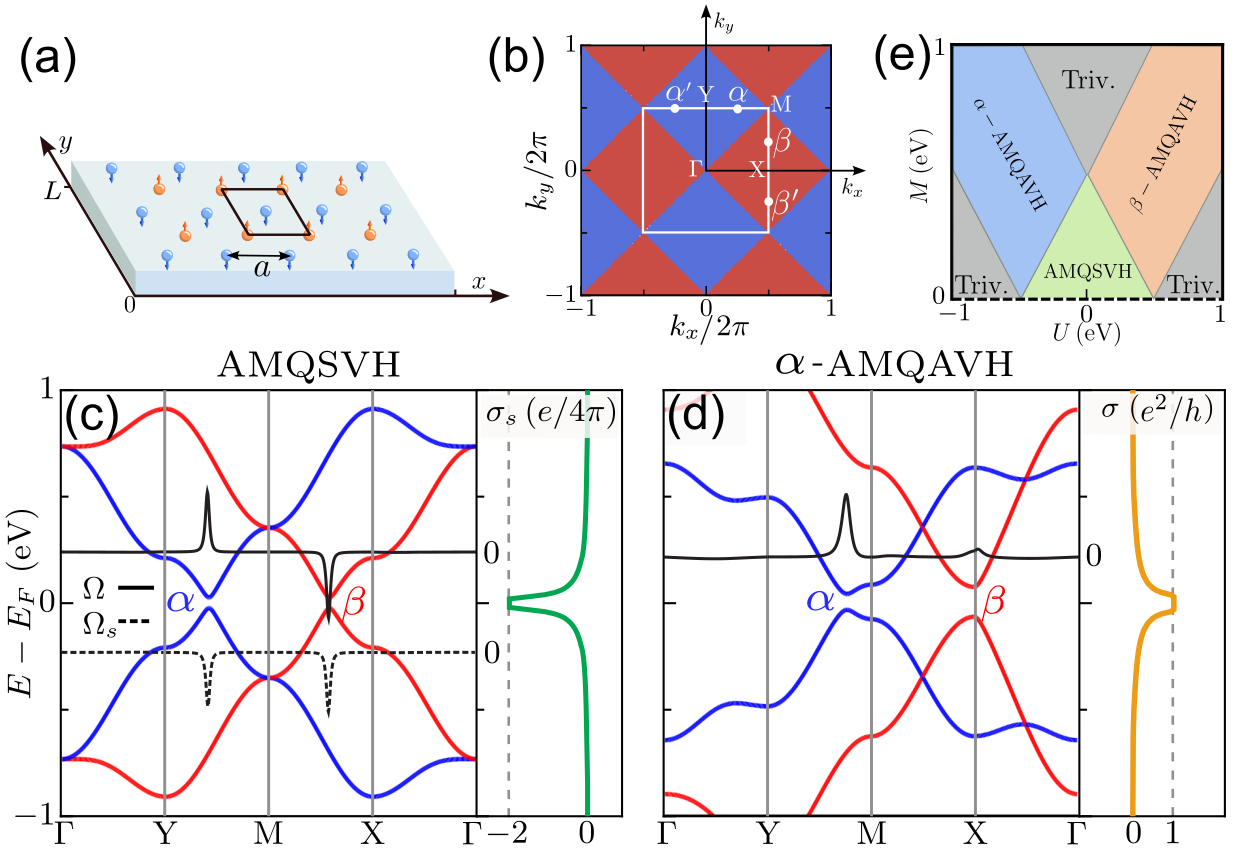}
	\caption{(a) 2D square-lattice AM model for Eq.~\ref{eq:tb}; the black square denotes the unit cell with lattice constant $a$. (b) Calculated $d$-wave AM spin texture in the first Brillouin zone, with valleys labeled $\alpha$ ($\alpha^{\prime}$) and $\beta$ ($\beta^{\prime}$). (c) AMQSVH phase: spin-resolved bands (red/blue), Berry curvature $\Omega$ (black solid), spin Berry curvature $\Omega_{s}$ (black dashed), and the resulting quantized spin Hall conductivity $\sigma_s$. (d) Same as (c) but for the $\alpha$-AMQAVH phase at $U<0$, showing a quantized Hall conductivity $\sigma$. (e) Phase diagram as a function of $U$ and $M$. Parameters are given in the main text and SM~\cite{SM}}.
	\label{fig:model}
\end{figure}

We demonstrate our framework through an effective tight-binding model based on a 2D square AFM lattice as in Fig.~\ref{fig:model}(a), which is written as:
\begin{align}\label{eq:tb}
	H = &\sum_{i,j} \left[ f_{i}^{\boldsymbol{\eta}_j} c_{i}^{\dagger}c_{i+\boldsymbol{\eta}_j} + g_{i}^{\boldsymbol{\kappa}_j} c_{i}^{\dagger}c_{i+\boldsymbol{\kappa}_j}+\mathrm{i}\lambda_{i}^{\boldsymbol{\eta}_j} c_{i}^{\dagger}\sigma_z c_{i+\boldsymbol{\eta}_j} + h.c. \right]\notag\\
	    &\qquad + M_{A,B}\sum_{i\in A,B} c_{i}^{\dagger}\sigma_z c_{i}+ U_{A,B}\sum_{i} c_{i}^{\dagger} c_{i}.
\end{align}
Here, $c_i^{\dagger}\left(c_i\right)$ are electron creation (annihilation) operators at site $i$ and $\boldsymbol{\sigma}$ is the Pauli matrix. The first and second terms describe hoppings between first (NN) and second (2NN) nearest-neighbors with hopping parameters $f_i^{\boldsymbol{\eta}_j}$ and $g_i^{\boldsymbol{\kappa}_j}$, where $\boldsymbol{\eta_j}$ and $\boldsymbol{\kappa}_j$ are the vectors connecting site $i$ to its NN and 2NN sites. The third term describes the SOC with strength $\lambda_i^{\boldsymbol{\eta}_j}$. The fourth term indicates AFM exchange fields on sublattices $A$ and $B$ with $M_A=-M_B=M$; The last term, describes the staggered potential with $U_A=-U_B=U$. Unless stated otherwise, we take M = 0.35 eV and $\mid$$U$$\mid$ = 0.4 eV throughout when these terms are included. All other parameters and model details are provided in the Supplemental Material (SM)~\cite{SM}.

Altermagnetism in this model is captured by sublattice-inequivalent hoppings together with the AFM exchange field~\cite{Zhu2025:SCPMA,Zhu2025:arXiv2,Liu2025:arxiv}, as confirmed by the calculated d-wave spin-polarization texture in Fig.~\ref{fig:model}(b). These terms also produce spin–valley–resolved Weyl pointsat the $\alpha$ ($\alpha'$) and $\beta$ ($\beta'$) valleys, consistent with previous studies~\cite{Antonenko2025:PRL,Wan2025:PRB}. Here $\alpha$ ($\beta$) and $\alpha'$ ($\beta'$) are related by mirror symmetry and exhibit identical spin splitting and topological character, as detailed in the SM~\cite{SM}. For brevity, we henceforth refer to them collectively as the $\alpha$ and $\beta$ valleys. These spin–valley–locked Weyl features, faithfully reproduced by our model, serve as the precursor to the AMQSVH and AMQAVH phases proposed below. Importantly, similar AM Weyl textures arise in realistic materials such as monolayers CrO~\cite{Guo2023:NCM,Chen2023:APL}, Nb$_2$SeTeO~\cite{Feng2025:arXiv}, Fe$_2$Te$_2$O~\cite{Zhang2025:arXiv}, and their derivatives~\cite{Xu2025:NSR}, underscoring the generality and material relevance of our theoretical framework.


\begin{figure}[t]
	\includegraphics[width=0.9\linewidth]{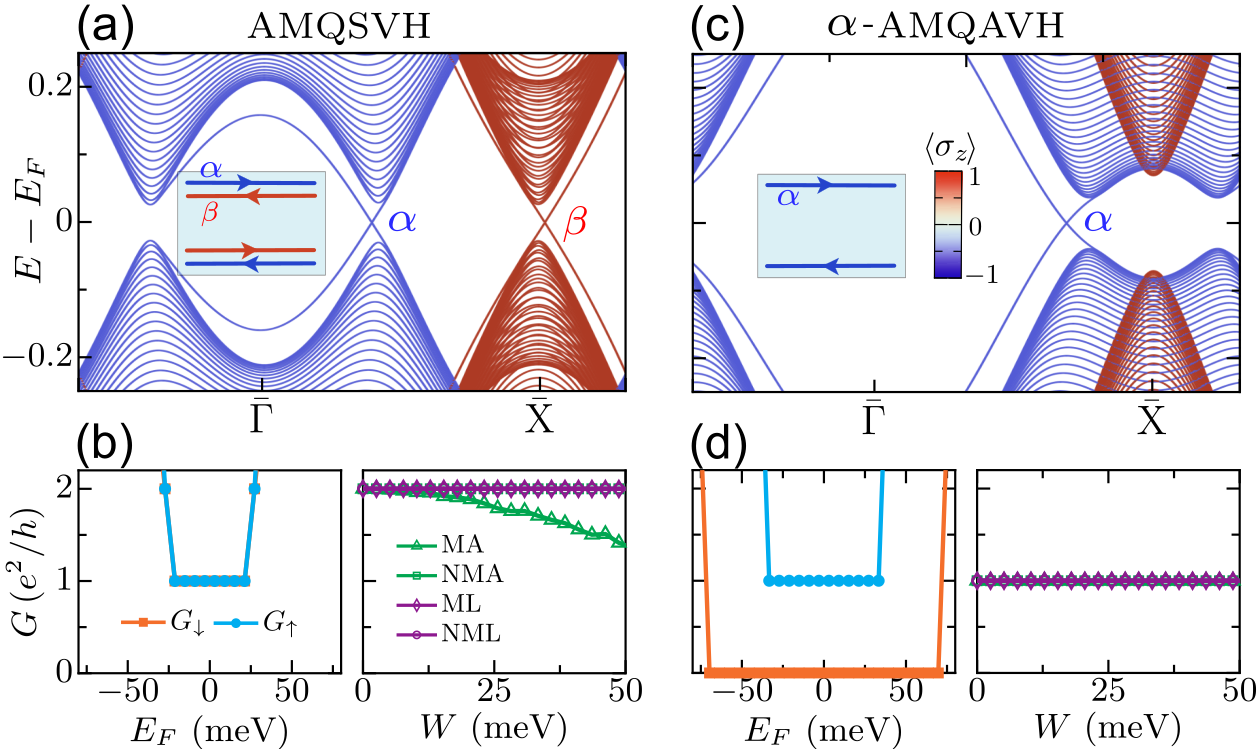}
	\caption{(a) Nanoribbon bands of the AMQSVH phase, showing helical SVML edge states located at $\alpha$ and $\beta$ valleys. (b) Spin-resolved conductance for (a) as a function of Fermi energy $E_F$ and disorder strength $W$, comparing magnetic Anderson (MA), nonmagnetic Anderson (NMA), magnetic long-range (ML), and nonmagnetic long-range (NML) disorder. (c,d) Same as (a,b) but for the $\alpha$-AMQAVH phase with a single chiral edge channel. Parameters taken from Fig.~\ref{fig:model}.}
	\label{fig:bands_transport}
\end{figure}

When SOC is introduced, each Weyl point is gapped as shown in Fig.~\ref{fig:model}(d). The resulting Berry curvature $\Omega(\mathbf{k})$ becomes strongly concentrated near $\alpha$ and $\beta$ valleys with opposite signs, yielding a vanishing total Chern number but a quantized valley Chern number $C_{\mathrm{v}}= C_{\alpha}-C_{\beta}=2$~\cite{Zhang2013:PNAS,Ren2016:RPP}, where $C_{\alpha} (C_{\beta})$ is the local topological charge around $\alpha$ ($\beta$) valley. Because spin-up (spin-down) bands are confined to the $\alpha$ ($\beta$) valley, spin mixing is strongly suppressed allowing spin to remain a good quantum number. The corresponding spin Berry curvature $\Omega_s(\mathbf{k})$ integrates to a fully quantized spin Hall conductivity $\sigma_{s}=-e/2\pi$, characterized by a spin Chern number~\cite{Sheng2006:PRL,Zhou2015:NL,Zhou2016:PRB,Rafael2025:PRB}, $C_{\mathrm{s}}=C^{\uparrow}-C^{\downarrow}=-2$. The coexistence of nonzero $C_{\mathrm{s}}$ and $C_{\mathrm{v}}$ reveals a new class of spin-valley-resolved QSH states endowed with dual topological protection.

To elucidate the topology of this phase, we examine the low-energy form of Eq.~\eqref{eq:tb}. Expanding the Hamiltonian near each valley yields a continuum model
\begin{equation}\label{eq:kp_model}
	H = \hbar v_{F} (k_x\tau_x + k_y\tau_y) + \gamma \lambda \tau_z,
\end{equation}
where $v_F$ is the Fermi velocity, $\gamma=\pm1$ labels the spin-valley sector, $\lambda$ is the SOC strength, and $\boldsymbol{\tau}$ denotes Pauli matrices acting on the conduction and valence bands. Integrating the spin-valley-resolved $\Omega(\mathbf{k})$ around each spin-valley pocket gives topological charges $(C_{\alpha}^{\downarrow},C_{\beta}^{\uparrow})=(1,-1)$, from which $C_{\mathrm{s}}= C_{\mathrm{v}}=2$. Since either index alone cannot fully capture the joint topology, we define a composite spin-valley Chern number
\begin{equation}\label{eq:sv-chernnumber}
	C_{\mathrm{sv}}=C_{\alpha}^{\downarrow} - C_{\beta}^{\uparrow},
\end{equation}
which uniquely characterizes the spin-valley-locked band topology. For this system $C_{\mathrm{sv}}=2$, implying a pair of helical SVML edge states. The calculated nanoribbon spectra in Fig.~\ref{fig:bands_transport}(a) and wave function probabilities in SM~\cite{SM} show that spin-up (spin-down) edge states are indeed bound to the $\alpha$ ($\beta$) valley, directly confirming the helical SVML AMQSVH states. This dual locking suppresses spin mixing and yields a genuine realization of the long-sought ideal QSH effect, verified by transport simulations showing robust quantized spin Hall conductance [Fig.~\ref{fig:bands_transport}(c)]. Unlike conventional QSH or QVH systems, which rely solely on spin- or valley-momentum locking~\cite{Hasan2010:RMP,Ren2016:RPP}, the AMQSVH phase combines both and can offer superior robustness against disorder.

Beyond the dual SVML protection, altermagnets provide further key advantage: the two counterpropagating SVML edge branches are linked by CRS~\cite{Ma2021:NC,Jungwirth2026:Nature}. As a result, breaking this symmetry can gap out one branch and convert the helical AMQSVH phase into a chiral AMQAVH phase, without introducing ferromagnetism or relying on finely tuned Chern-gap reopening as in conventional schemes~\cite{Yu2010:Science,Chang2013:Science,Chang2023:RMP}. We realize this symmetry-controlled transition through $U$, a generic rotation-symmetry–breaking perturbation widely used for band-structure engineering and topological phase manipulation in two-sublattice 2D materials, and readily tunable by an external electric field~\cite{schaibley2016:NRM,Li2016:NN,Li2018:Science,Ren2016:RPP,Zhou2021:PRL}. The $U$-dependent band structures in [Fig.~\ref{fig:model}(d) and the phase diagram in [Fig.~\ref{fig:model}(e) (see more details in SM [58]) show that for $U<0$ the $\beta$-valley gap closes and reopens in a trivial form, while the $\alpha$-valley gap remains nontrivial, yielding an $\alpha$-AMQAVH phase as corroborated by the Berry-curvature distribution and the quantized Hall conductance in Fig.~\ref{fig:model}(d). Correspondingly, the helical edge pair collapses into a single chiral SVML channel with fixed spin–valley polarization (spin down in the $\alpha$ valley), producing disorder-robust chiral transport [Fig.~\ref{fig:bands_transport}(c-d)]. Reversing the sign of $U$ instead selects the $\beta$-AMQAVH phase with the opposite spin–valley polarization (see SM~\cite{SM}). Because both the magnitude and sign of $U$ can be continuously tuned by gating~\cite{schaibley2016:NRM,Li2016:NN,Li2018:Science,Ren2016:RPP,Zhou2021:PRL}, these results realize an all-electrical switch between helical and chiral topological transport, positioning altermagnets as a scalable platform for electrically reconfigurable spin–valley topological devices.

A defining hallmark of topological phases is their disorder-resilient transport. The robustness of quantized conductance under specific types of disorder provides both a fingerprint of the underlying topology and a criterion for device utility~\cite{Hasan2010:RMP,Qi2011:RMP,Ren2016:RPP}. For instance, conventional QSH states protected by TRS remain dissipationless under nonmagnetic disorder but are destroyed by magnetic impurities that break TRS and disrupt spin–momentum locking~\cite{Hasan2010:RMP,Qi2011:RMP}. Similarly, quantum valley Hall states are robust against long-range disorder yet fragile under short-range disorder that induces intervalley scattering and lifts valley–momentum locking~\cite{Zhou2021:PRL,Ren2016:RPP}. To evaluate the disorder robustness of the AMQSVH state, we calculate spin-resolved conductance using the Landauer–Büttiker formalism and Green’s-function method~\cite{Sancho1984:JPFMP,Jiang2009:PRB}. We consider both magnetic and nonmagnetic disorder—distinguished by their ability to flip spins—and short- versus long-range correlation lengths, $\zeta$, relative to the lattice constant $a$. Conductance calculations for AMQSVH nanoribbons under four representative disorder types (magnetic/nonmagnetic Anderson, $\zeta$=0; magnetic/nonmagnetic long-range, $\zeta$=5$a$) are shown in Fig.~\ref{fig:bands_transport}(c), and more details are given in the SM~\cite{SM}. Consistent with its dual protection, the AMQSVH phase retains quantized conductance under both short- and long-range nonmagnetic disorder, similar to conventional QSH systems. However, unlike QSH phases that are easily destroyed by long-range magnetic fluctuations~\cite{Konig2007:Science,Qi2011:RMP}, AMQSVH remains quantized because its additional valley–momentum locking protection survives. Only magnetic Anderson disorder—capable of simultaneous spin flips and intervalley scattering—breaks the SVML condition and suppresses quantization. These results directly confirm that dual SVML protection is responsible for the enhanced robustness of the AMQSVH state. To overcome this remaining fragility against magnetic Anderson disorder, one can selectively gap a single valley to remove one helical branch, driving the system into the chiral AMQAVH phase discussed above. Unlike conventional QAH states where spin and valley indices are ill-defined, AMQAVH preserves SVML character and exhibits quantized spin–valley–polarized chiral transport robust against all disorder types [Fig.~\ref{fig:bands_transport}(d)].

\begin{figure}[t!]
	\includegraphics[width=\linewidth]{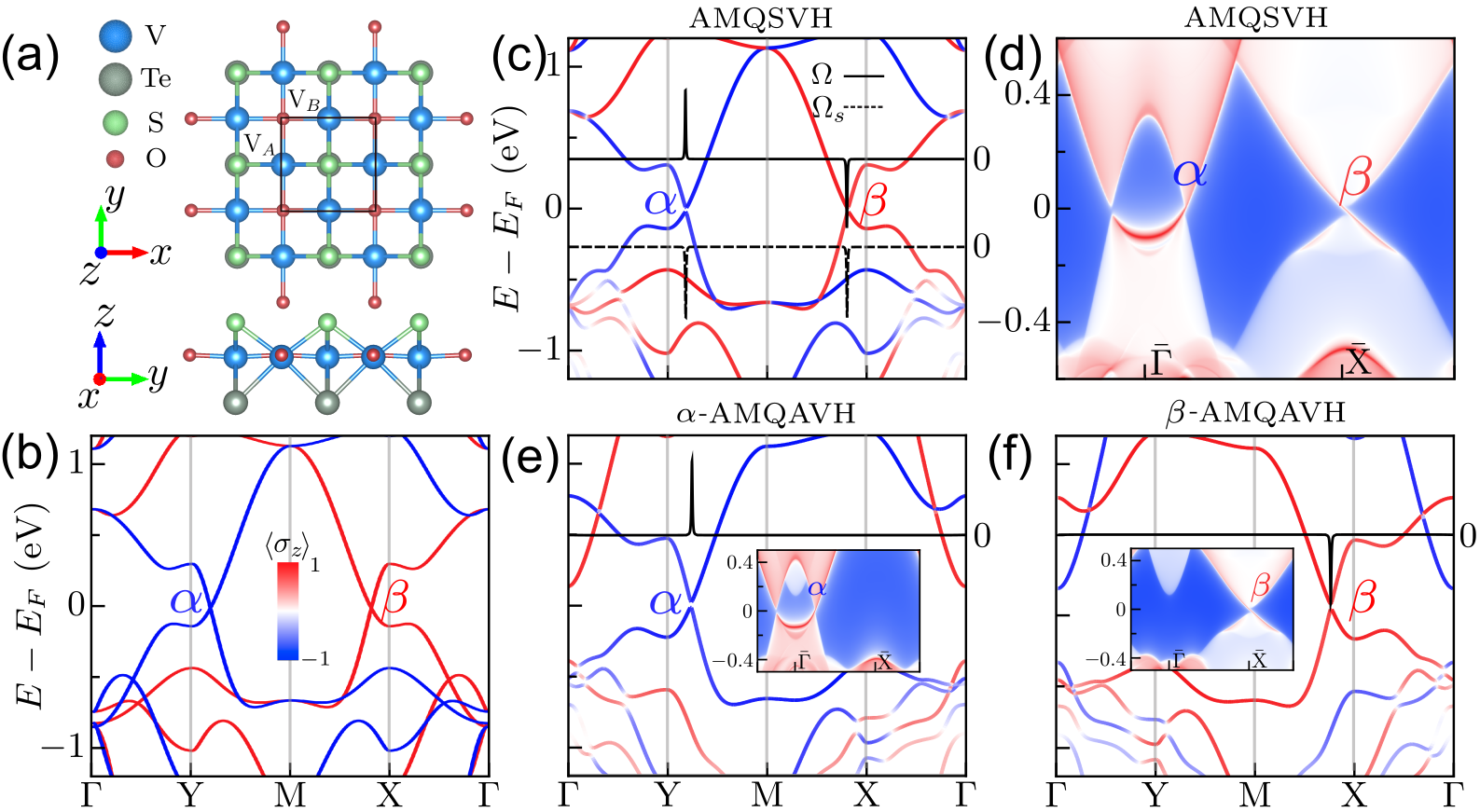}
	\caption{(a) Top and side views of monolayer V$_2$TeSO. (b) Spin-resolved bands of (a) without SOC, showing AM Weyl points. (c) Same as (b) but with SOC; the corresponding (spin) Berry curvatures indicate the AMQSVH phase, whose helical edge states are shown in (d). (e) Same as (c) but for Ti substitution on the $V_A$ sublattice, which induces $U<0$ and yields the $\alpha$-AMQAVH phase with chiral edge state (inset). (f) Same as (e) but for Ti substitution on the $V_B$ sublattice ($U>0$), yielding the $\beta$-AMQAVH state.}
	\label{fig:dft_realization}
\end{figure}

Experimentally, our proposed AMQSVH and AMQAVH phases, together with their electrically controlled interconversion, are well supported by realistic material platforms. Guided by the symmetry criteria and model insights developed above, our first-principles calculations identify monolayers V$_2$STeO, VO, and related derivatives as AMQSVH candidates (see SM~\cite{SM}). As a representative example, we consider monolayer V$_2$STeO to illustrate a concrete realization: without SOC its spectrum hosts spin–valley–locked Weyl points [Fig.~\ref{fig:dft_realization}(b)], whereas SOC opens gaps of order 20 meV at both valleys [Fig.~\ref{fig:dft_realization}(c)], producing the targeted AMQSVH phase. The corresponding (spin) Berry-curvature distributions and Wannier-based edge-state calculations, which reveal helical SVML edge modes [Figs.~\ref{fig:dft_realization}(c,d)], further confirm this identification.

To realize the chiral AMQAVH state, we introduce a controllable sublattice asymmetry that generates an effective $U$. Concretely, we implement this via alloy engineering, replacing the $V$ atom on one sublattice (V$_A$) by Ti, a standard strategy for engineering sublattice potentials and tuning band topology~\cite{schaibley2016:NRM,Li2016:NN,Li2018:Science,Ren2016:RPP,Zhou2021:PRL}. This substitution selectively trivializes the $\beta$-valley gap while preserving a topological gap at $\alpha$ valley, yielding the $\alpha$-AMQAVH phase as confirmed by the Berry curvature and the emergence of a single chiral SVML edge channel [Fig.~\ref{fig:dft_realization}(e)]. Replacing instead the V atom on the opposite sublattice (V$_B$) reverses the effective sign of $U$ and switches the system to the $\beta$-AMQAVH phase [Fig.~\ref{fig:dft_realization}(f)], directly validating the symmetry-based design principle. This material route to AMQAVH state is not specific to V$_2$STeO and can extend to other monolayer families (e.g. Nb$_2$SeTeO and VO) as seen in SM~\cite{SM}, underscoring the universality and experimental feasibility of our proposal.

By integrating three defining ingredients of altermagnets—valley-locked spin splitting, rotation-symmetry–connected valleys, and intrinsic antiferromagnetic order—we establish a unified and symmetry-driven framework for realizing and electrically interconverting helical and chiral topological transport within a single material platform. The resulting SVML edge channels provide dual topological protection, yielding enhanced robustness against disorder while enabling deterministic, electrical selection of the transmitted channel type and its spin–valley polarization. More broadly, because our proposed mechanism is rooted in symmetry and band topology rather than material-specific chemistry, it is readily extendable to topological photonic~\cite{Ozawa2019:RMP}, phononic~\cite{zhu2023:RPP}, magnonic~\cite{Chen2025:Nature}, and superconducting systems~\cite{Zhu2025:arXiv,Zhou2020:PRL, Amundsen2024:RMP}. This scalability opens a pathway toward low-dissipation, electrically programmable topological circuitry and multifunctional quantum devices~\cite{vZutic2004:RMP}, bridging condensed-matter physics, materials science, and emerging quantum technologies.

\textit{Acknowledgments--}
This work is supported by the Zhejiang Provincial Natural Science Foundation of China (LR25A040001), the National Natural Science Foundation of China (12474155 and 11904250). The computational resources for this research were provided by the High Performance Computing Platform at the Eastern Institute of Technology, Ningbo.

X.C., J.Z., and B.H. contributed equally to this work.

\bibliography{AMTOP_arXiv_Refs}{}

\begin{thebibliography}{78}%
\makeatletter
\providecommand \@ifxundefined [1]{%
 \@ifx{#1\undefined}
}%
\providecommand \@ifnum [1]{%
 \ifnum #1\expandafter \@firstoftwo
 \else \expandafter \@secondoftwo
 \fi
}%
\providecommand \@ifx [1]{%
 \ifx #1\expandafter \@firstoftwo
 \else \expandafter \@secondoftwo
 \fi
}%
\providecommand \natexlab [1]{#1}%
\providecommand \enquote  [1]{``#1''}%
\providecommand \bibnamefont  [1]{#1}%
\providecommand \bibfnamefont [1]{#1}%
\providecommand \citenamefont [1]{#1}%
\providecommand \href@noop [0]{\@secondoftwo}%
\providecommand \href [0]{\begingroup \@sanitize@url \@href}%
\providecommand \@href[1]{\@@startlink{#1}\@@href}%
\providecommand \@@href[1]{\endgroup#1\@@endlink}%
\providecommand \@sanitize@url [0]{\catcode `\\12\catcode `\$12\catcode
  `\&12\catcode `\#12\catcode `\^12\catcode `\_12\catcode `\%12\relax}%
\providecommand \@@startlink[1]{}%
\providecommand \@@endlink[0]{}%
\providecommand \url  [0]{\begingroup\@sanitize@url \@url }%
\providecommand \@url [1]{\endgroup\@href {#1}{\urlprefix }}%
\providecommand \urlprefix  [0]{URL }%
\providecommand \Eprint [0]{\href }%
\providecommand \doibase [0]{https://doi.org/}%
\providecommand \selectlanguage [0]{\@gobble}%
\providecommand \bibinfo  [0]{\@secondoftwo}%
\providecommand \bibfield  [0]{\@secondoftwo}%
\providecommand \translation [1]{[#1]}%
\providecommand \BibitemOpen [0]{}%
\providecommand \bibitemStop [0]{}%
\providecommand \bibitemNoStop [0]{.\EOS\space}%
\providecommand \EOS [0]{\spacefactor3000\relax}%
\providecommand \BibitemShut  [1]{\csname bibitem#1\endcsname}%
\let\auto@bib@innerbib\@empty
\bibitem [{\citenamefont {Hasan}\ and\ \citenamefont
  {Kane}(2010)}]{Hasan2010:RMP}%
  \BibitemOpen
  \bibfield  {author} {\bibinfo {author} {\bibfnamefont {M.~Z.}\ \bibnamefont
  {Hasan}}\ and\ \bibinfo {author} {\bibfnamefont {C.~L.}\ \bibnamefont
  {Kane}},\ }\bibfield  {title} {\bibinfo {title} {Colloquium: {{Topological}}
  insulators},\ }\href {https://doi.org/10.1103/RevModPhys.82.3045} {\bibfield
  {journal} {\bibinfo  {journal} {Rev. Mod. Phys.}\ }\textbf {\bibinfo {volume}
  {82}},\ \bibinfo {pages} {3045} (\bibinfo {year} {2010})}\BibitemShut
  {NoStop}%
\bibitem [{\citenamefont {Qi}\ and\ \citenamefont {Zhang}(2011)}]{Qi2011:RMP}%
  \BibitemOpen
  \bibfield  {author} {\bibinfo {author} {\bibfnamefont {X.-L.}\ \bibnamefont
  {Qi}}\ and\ \bibinfo {author} {\bibfnamefont {S.-C.}\ \bibnamefont {Zhang}},\
  }\bibfield  {title} {\bibinfo {title} {Topological insulators and
  superconductors},\ }\href {https://doi.org/10.1103/RevModPhys.83.1057}
  {\bibfield  {journal} {\bibinfo  {journal} {Rev. Mod. Phys.}\ }\textbf
  {\bibinfo {volume} {83}},\ \bibinfo {pages} {1057} (\bibinfo {year}
  {2011})}\BibitemShut {NoStop}%
\bibitem [{\citenamefont {Matusalem}\ \emph {et~al.}(2019)\citenamefont
  {Matusalem}, \citenamefont {Marques}, \citenamefont {Teles}, \citenamefont
  {Matthes}, \citenamefont {Furthm{\"u}ller},\ and\ \citenamefont
  {Bechstedt}}]{Matusalem2019:PRB}%
  \BibitemOpen
  \bibfield  {author} {\bibinfo {author} {\bibfnamefont {F.}~\bibnamefont
  {Matusalem}}, \bibinfo {author} {\bibfnamefont {M.}~\bibnamefont {Marques}},
  \bibinfo {author} {\bibfnamefont {L.~K.}\ \bibnamefont {Teles}}, \bibinfo
  {author} {\bibfnamefont {L.}~\bibnamefont {Matthes}}, \bibinfo {author}
  {\bibfnamefont {J.}~\bibnamefont {Furthm{\"u}ller}},\ and\ \bibinfo {author}
  {\bibfnamefont {F.}~\bibnamefont {Bechstedt}},\ }\bibfield  {title} {\bibinfo
  {title} {Quantization of spin {{Hall}} conductivity in two-dimensional
  topological insulators versus symmetry and spin-orbit interaction},\ }\href
  {https://doi.org/10.1103/PhysRevB.100.245430} {\bibfield  {journal} {\bibinfo
   {journal} {Phys. Rev. B}\ }\textbf {\bibinfo {volume} {100}},\ \bibinfo
  {pages} {245430} (\bibinfo {year} {2019})}\BibitemShut {NoStop}%
\bibitem [{\citenamefont {Monaco}\ and\ \citenamefont {Ul{\v
  c}akar}(2020)}]{Monaco2020:PRB}%
  \BibitemOpen
  \bibfield  {author} {\bibinfo {author} {\bibfnamefont {D.}~\bibnamefont
  {Monaco}}\ and\ \bibinfo {author} {\bibfnamefont {L.}~\bibnamefont {Ul{\v
  c}akar}},\ }\bibfield  {title} {\bibinfo {title} {Spin {{Hall}} conductivity
  in insulators with nonconserved spin},\ }\href
  {https://doi.org/10.1103/PhysRevB.102.125138} {\bibfield  {journal} {\bibinfo
   {journal} {Phys. Rev. B}\ }\textbf {\bibinfo {volume} {102}},\ \bibinfo
  {pages} {125138} (\bibinfo {year} {2020})}\BibitemShut {NoStop}%
\bibitem [{\citenamefont {Maciejko}\ \emph {et~al.}(2009)\citenamefont
  {Maciejko}, \citenamefont {Liu}, \citenamefont {Oreg}, \citenamefont {Qi},
  \citenamefont {Wu},\ and\ \citenamefont {Zhang}}]{Maciejko2009:PRL}%
  \BibitemOpen
  \bibfield  {author} {\bibinfo {author} {\bibfnamefont {J.}~\bibnamefont
  {Maciejko}}, \bibinfo {author} {\bibfnamefont {C.}~\bibnamefont {Liu}},
  \bibinfo {author} {\bibfnamefont {Y.}~\bibnamefont {Oreg}}, \bibinfo {author}
  {\bibfnamefont {X.-L.}\ \bibnamefont {Qi}}, \bibinfo {author} {\bibfnamefont
  {C.}~\bibnamefont {Wu}},\ and\ \bibinfo {author} {\bibfnamefont {S.-C.}\
  \bibnamefont {Zhang}},\ }\bibfield  {title} {\bibinfo {title} {Kondo
  {{Effect}} in the {{Helical Edge Liquid}} of the {{Quantum Spin Hall
  State}}},\ }\href {https://doi.org/10.1103/PhysRevLett.102.256803} {\bibfield
   {journal} {\bibinfo  {journal} {Phys. Rev. Lett.}\ }\textbf {\bibinfo
  {volume} {102}},\ \bibinfo {pages} {256803} (\bibinfo {year}
  {2009})}\BibitemShut {NoStop}%
\bibitem [{\citenamefont {Kimme}\ \emph {et~al.}(2016)\citenamefont {Kimme},
  \citenamefont {Rosenow},\ and\ \citenamefont {Brataas}}]{Kimme2016:PRB}%
  \BibitemOpen
  \bibfield  {author} {\bibinfo {author} {\bibfnamefont {L.}~\bibnamefont
  {Kimme}}, \bibinfo {author} {\bibfnamefont {B.}~\bibnamefont {Rosenow}},\
  and\ \bibinfo {author} {\bibfnamefont {A.}~\bibnamefont {Brataas}},\
  }\bibfield  {title} {\bibinfo {title} {Backscattering in helical edge states
  from a magnetic impurity and {{Rashba}} disorder},\ }\href
  {https://doi.org/10.1103/PhysRevB.93.081301} {\bibfield  {journal} {\bibinfo
  {journal} {Phys. Rev. B}\ }\textbf {\bibinfo {volume} {93}},\ \bibinfo
  {pages} {081301} (\bibinfo {year} {2016})}\BibitemShut {NoStop}%
\bibitem [{\citenamefont {J{\"a}ck}\ \emph {et~al.}(2020)\citenamefont
  {J{\"a}ck}, \citenamefont {Xie}, \citenamefont {Andrei~Bernevig},\ and\
  \citenamefont {Yazdani}}]{Jack2020:PNAS}%
  \BibitemOpen
  \bibfield  {author} {\bibinfo {author} {\bibfnamefont {B.}~\bibnamefont
  {J{\"a}ck}}, \bibinfo {author} {\bibfnamefont {Y.}~\bibnamefont {Xie}},
  \bibinfo {author} {\bibfnamefont {B.}~\bibnamefont {Andrei~Bernevig}},\ and\
  \bibinfo {author} {\bibfnamefont {A.}~\bibnamefont {Yazdani}},\ }\bibfield
  {title} {\bibinfo {title} {Observation of backscattering induced by magnetism
  in a topological edge state},\ }\href
  {https://doi.org/10.1073/pnas.2005071117} {\bibfield  {journal} {\bibinfo
  {journal} {Proc. Natl. Acad. Sci.}\ }\textbf {\bibinfo {volume} {117}},\
  \bibinfo {pages} {16214} (\bibinfo {year} {2020})}\BibitemShut {NoStop}%
\bibitem [{\citenamefont {Zhou}\ \emph {et~al.}(2021)\citenamefont {Zhou},
  \citenamefont {Cheng}, \citenamefont {Schleenvoigt}, \citenamefont
  {Sch\"uffelgen}, \citenamefont {Jiang}, \citenamefont {Yang},\ and\
  \citenamefont {\ifmmode \check{Z}\else \v{Z}\fi{}uti\ifmmode~\acute{c}\else
  \'{c}\fi{}}}]{Zhou2021:PRL}%
  \BibitemOpen
  \bibfield  {author} {\bibinfo {author} {\bibfnamefont {T.}~\bibnamefont
  {Zhou}}, \bibinfo {author} {\bibfnamefont {S.}~\bibnamefont {Cheng}},
  \bibinfo {author} {\bibfnamefont {M.}~\bibnamefont {Schleenvoigt}}, \bibinfo
  {author} {\bibfnamefont {P.}~\bibnamefont {Sch\"uffelgen}}, \bibinfo {author}
  {\bibfnamefont {H.}~\bibnamefont {Jiang}}, \bibinfo {author} {\bibfnamefont
  {Z.}~\bibnamefont {Yang}},\ and\ \bibinfo {author} {\bibfnamefont
  {I.}~\bibnamefont {\ifmmode \check{Z}\else
  \v{Z}\fi{}uti\ifmmode~\acute{c}\else \'{c}\fi{}}},\ }\bibfield  {title}
  {\bibinfo {title} {Quantum {{Spin-Valley Hall Kink States}}: {{From Concept}}
  to {{Materials Design}}},\ }\href
  {https://doi.org/10.1103/PhysRevLett.127.116402} {\bibfield  {journal}
  {\bibinfo  {journal} {Phys. Rev. Lett.}\ }\textbf {\bibinfo {volume} {127}},\
  \bibinfo {pages} {116402} (\bibinfo {year} {2021})}\BibitemShut {NoStop}%
\bibitem [{\citenamefont {Yu}\ \emph {et~al.}(2010)\citenamefont {Yu},
  \citenamefont {Zhang}, \citenamefont {Zhang}, \citenamefont {Zhang},
  \citenamefont {Dai},\ and\ \citenamefont {Fang}}]{Yu2010:Science}%
  \BibitemOpen
  \bibfield  {author} {\bibinfo {author} {\bibfnamefont {R.}~\bibnamefont
  {Yu}}, \bibinfo {author} {\bibfnamefont {W.}~\bibnamefont {Zhang}}, \bibinfo
  {author} {\bibfnamefont {H.-J.}\ \bibnamefont {Zhang}}, \bibinfo {author}
  {\bibfnamefont {S.-C.}\ \bibnamefont {Zhang}}, \bibinfo {author}
  {\bibfnamefont {X.}~\bibnamefont {Dai}},\ and\ \bibinfo {author}
  {\bibfnamefont {Z.}~\bibnamefont {Fang}},\ }\bibfield  {title} {\bibinfo
  {title} {Quantized {{Anomalous Hall Effect}} in {{Magnetic Topological
  Insulators}}},\ }\href {https://doi.org/10.1126/science.1187485} {\bibfield
  {journal} {\bibinfo  {journal} {Science}\ }\textbf {\bibinfo {volume}
  {329}},\ \bibinfo {pages} {61} (\bibinfo {year} {2010})}\BibitemShut
  {NoStop}%
\bibitem [{\citenamefont {Chang}\ \emph {et~al.}(2013)\citenamefont {Chang},
  \citenamefont {Zhang}, \citenamefont {Feng}, \citenamefont {Shen},
  \citenamefont {Zhang}, \citenamefont {Guo}, \citenamefont {Li}, \citenamefont
  {Ou}, \citenamefont {Wei}, \citenamefont {Wang}, \citenamefont {Ji},
  \citenamefont {Feng}, \citenamefont {Ji}, \citenamefont {Chen}, \citenamefont
  {Jia}, \citenamefont {Dai}, \citenamefont {Fang}, \citenamefont {Zhang},
  \citenamefont {He}, \citenamefont {Wang}, \citenamefont {Lu}, \citenamefont
  {Ma},\ and\ \citenamefont {Xue}}]{Chang2013:Science}%
  \BibitemOpen
  \bibfield  {author} {\bibinfo {author} {\bibfnamefont {C.-Z.}\ \bibnamefont
  {Chang}}, \bibinfo {author} {\bibfnamefont {J.}~\bibnamefont {Zhang}},
  \bibinfo {author} {\bibfnamefont {X.}~\bibnamefont {Feng}}, \bibinfo {author}
  {\bibfnamefont {J.}~\bibnamefont {Shen}}, \bibinfo {author} {\bibfnamefont
  {Z.}~\bibnamefont {Zhang}}, \bibinfo {author} {\bibfnamefont
  {M.}~\bibnamefont {Guo}}, \bibinfo {author} {\bibfnamefont {K.}~\bibnamefont
  {Li}}, \bibinfo {author} {\bibfnamefont {Y.}~\bibnamefont {Ou}}, \bibinfo
  {author} {\bibfnamefont {P.}~\bibnamefont {Wei}}, \bibinfo {author}
  {\bibfnamefont {L.-L.}\ \bibnamefont {Wang}}, \bibinfo {author}
  {\bibfnamefont {Z.-Q.}\ \bibnamefont {Ji}}, \bibinfo {author} {\bibfnamefont
  {Y.}~\bibnamefont {Feng}}, \bibinfo {author} {\bibfnamefont {S.}~\bibnamefont
  {Ji}}, \bibinfo {author} {\bibfnamefont {X.}~\bibnamefont {Chen}}, \bibinfo
  {author} {\bibfnamefont {J.}~\bibnamefont {Jia}}, \bibinfo {author}
  {\bibfnamefont {X.}~\bibnamefont {Dai}}, \bibinfo {author} {\bibfnamefont
  {Z.}~\bibnamefont {Fang}}, \bibinfo {author} {\bibfnamefont {S.-C.}\
  \bibnamefont {Zhang}}, \bibinfo {author} {\bibfnamefont {K.}~\bibnamefont
  {He}}, \bibinfo {author} {\bibfnamefont {Y.}~\bibnamefont {Wang}}, \bibinfo
  {author} {\bibfnamefont {L.}~\bibnamefont {Lu}}, \bibinfo {author}
  {\bibfnamefont {X.-C.}\ \bibnamefont {Ma}},\ and\ \bibinfo {author}
  {\bibfnamefont {Q.-K.}\ \bibnamefont {Xue}},\ }\bibfield  {title} {\bibinfo
  {title} {Experimental observation of the quantum anomalous hall effect in a
  magnetic topological insulator},\ }\href
  {https://doi.org/10.1126/science.1234414} {\bibfield  {journal} {\bibinfo
  {journal} {Science}\ }\textbf {\bibinfo {volume} {340}},\ \bibinfo {pages}
  {167} (\bibinfo {year} {2013})}\BibitemShut {NoStop}%
\bibitem [{\citenamefont {Chang}\ \emph {et~al.}(2023)\citenamefont {Chang},
  \citenamefont {Liu},\ and\ \citenamefont {MacDonald}}]{Chang2023:RMP}%
  \BibitemOpen
  \bibfield  {author} {\bibinfo {author} {\bibfnamefont {C.-Z.}\ \bibnamefont
  {Chang}}, \bibinfo {author} {\bibfnamefont {C.-X.}\ \bibnamefont {Liu}},\
  and\ \bibinfo {author} {\bibfnamefont {A.~H.}\ \bibnamefont {MacDonald}},\
  }\bibfield  {title} {\bibinfo {title} {Colloquium: {{Quantum}} anomalous
  {{Hall}} effect},\ }\href {https://doi.org/10.1103/RevModPhys.95.011002}
  {\bibfield  {journal} {\bibinfo  {journal} {Rev. Mod. Phys.}\ }\textbf
  {\bibinfo {volume} {95}},\ \bibinfo {pages} {011002} (\bibinfo {year}
  {2023})}\BibitemShut {NoStop}%
\bibitem [{\citenamefont {Tokura}\ \emph {et~al.}(2019)\citenamefont {Tokura},
  \citenamefont {Yasuda},\ and\ \citenamefont {Tsukazaki}}]{Tokura2019:NRP}%
  \BibitemOpen
  \bibfield  {author} {\bibinfo {author} {\bibfnamefont {Y.}~\bibnamefont
  {Tokura}}, \bibinfo {author} {\bibfnamefont {K.}~\bibnamefont {Yasuda}},\
  and\ \bibinfo {author} {\bibfnamefont {A.}~\bibnamefont {Tsukazaki}},\
  }\bibfield  {title} {\bibinfo {title} {Magnetic topological insulators},\
  }\href {https://doi.org/10.1038/s42254-018-0011-5} {\bibfield  {journal}
  {\bibinfo  {journal} {Nat. Rev. Phys.}\ }\textbf {\bibinfo {volume} {1}},\
  \bibinfo {pages} {126} (\bibinfo {year} {2019})}\BibitemShut {NoStop}%
\bibitem [{\citenamefont {Wu}\ \emph {et~al.}(2007)\citenamefont {Wu},
  \citenamefont {Sun}, \citenamefont {Fradkin},\ and\ \citenamefont
  {Zhang}}]{Wu2007:PRB}%
  \BibitemOpen
  \bibfield  {author} {\bibinfo {author} {\bibfnamefont {C.}~\bibnamefont
  {Wu}}, \bibinfo {author} {\bibfnamefont {K.}~\bibnamefont {Sun}}, \bibinfo
  {author} {\bibfnamefont {E.}~\bibnamefont {Fradkin}},\ and\ \bibinfo {author}
  {\bibfnamefont {S.-C.}\ \bibnamefont {Zhang}},\ }\bibfield  {title} {\bibinfo
  {title} {Fermi liquid instabilities in the spin channel},\ }\href
  {https://doi.org/10.1103/PhysRevB.75.115103} {\bibfield  {journal} {\bibinfo
  {journal} {Phys. Rev. B}\ }\textbf {\bibinfo {volume} {75}},\ \bibinfo
  {pages} {115103} (\bibinfo {year} {2007})}\BibitemShut {NoStop}%
\bibitem [{\citenamefont {Hayami}\ \emph {et~al.}(2019)\citenamefont {Hayami},
  \citenamefont {Yanagi},\ and\ \citenamefont {Kusunose}}]{Hayami2019:JPSJ}%
  \BibitemOpen
  \bibfield  {author} {\bibinfo {author} {\bibfnamefont {S.}~\bibnamefont
  {Hayami}}, \bibinfo {author} {\bibfnamefont {Y.}~\bibnamefont {Yanagi}},\
  and\ \bibinfo {author} {\bibfnamefont {H.}~\bibnamefont {Kusunose}},\
  }\bibfield  {title} {\bibinfo {title} {Momentum-{{Dependent Spin Splitting}}
  by {{Collinear Antiferromagnetic Ordering}}},\ }\href
  {https://doi.org/10.7566/JPSJ.88.123702} {\bibfield  {journal} {\bibinfo
  {journal} {J. Phys. Soc. Jpn.}\ }\textbf {\bibinfo {volume} {88}},\ \bibinfo
  {pages} {123702} (\bibinfo {year} {2019})}\BibitemShut {NoStop}%
\bibitem [{\citenamefont {Yuan}\ \emph {et~al.}(2020)\citenamefont {Yuan},
  \citenamefont {Wang}, \citenamefont {Luo}, \citenamefont {Rashba},\ and\
  \citenamefont {Zunger}}]{Yuan2020:PRB}%
  \BibitemOpen
  \bibfield  {author} {\bibinfo {author} {\bibfnamefont {L.-D.}\ \bibnamefont
  {Yuan}}, \bibinfo {author} {\bibfnamefont {Z.}~\bibnamefont {Wang}}, \bibinfo
  {author} {\bibfnamefont {J.-W.}\ \bibnamefont {Luo}}, \bibinfo {author}
  {\bibfnamefont {E.~I.}\ \bibnamefont {Rashba}},\ and\ \bibinfo {author}
  {\bibfnamefont {A.}~\bibnamefont {Zunger}},\ }\bibfield  {title} {\bibinfo
  {title} {Giant momentum-dependent spin splitting in centrosymmetric
  low-\${{Z}}\$ antiferromagnets},\ }\href
  {https://doi.org/10.1103/PhysRevB.102.014422} {\bibfield  {journal} {\bibinfo
   {journal} {Phys. Rev. B}\ }\textbf {\bibinfo {volume} {102}},\ \bibinfo
  {pages} {014422} (\bibinfo {year} {2020})}\BibitemShut {NoStop}%
\bibitem [{\citenamefont {{\v S}mejkal}\ \emph {et~al.}(2020)\citenamefont {{\v
  S}mejkal}, \citenamefont {{Gonz{\'a}lez-Hern{\'a}ndez}}, \citenamefont
  {Jungwirth},\ and\ \citenamefont {Sinova}}]{Smejkal2020:SA}%
  \BibitemOpen
  \bibfield  {author} {\bibinfo {author} {\bibfnamefont {L.}~\bibnamefont {{\v
  S}mejkal}}, \bibinfo {author} {\bibfnamefont {R.}~\bibnamefont
  {{Gonz{\'a}lez-Hern{\'a}ndez}}}, \bibinfo {author} {\bibfnamefont
  {T.}~\bibnamefont {Jungwirth}},\ and\ \bibinfo {author} {\bibfnamefont
  {J.}~\bibnamefont {Sinova}},\ }\bibfield  {title} {\bibinfo {title} {Crystal
  time-reversal symmetry breaking and spontaneous {{Hall}} effect in collinear
  antiferromagnets},\ }\href {https://doi.org/10.1126/sciadv.aaz8809}
  {\bibfield  {journal} {\bibinfo  {journal} {Sci. Adv.}\ }\textbf {\bibinfo
  {volume} {6}},\ \bibinfo {pages} {eaaz8809} (\bibinfo {year}
  {2020})}\BibitemShut {NoStop}%
\bibitem [{\citenamefont {Mazin}\ \emph {et~al.}(2021)\citenamefont {Mazin},
  \citenamefont {Koepernik}, \citenamefont {Johannes}, \citenamefont
  {{Gonz{\'a}lez-Hern{\'a}ndez}},\ and\ \citenamefont {{\v
  S}mejkal}}]{Mazin2021:PNAS}%
  \BibitemOpen
  \bibfield  {author} {\bibinfo {author} {\bibfnamefont {I.~I.}\ \bibnamefont
  {Mazin}}, \bibinfo {author} {\bibfnamefont {K.}~\bibnamefont {Koepernik}},
  \bibinfo {author} {\bibfnamefont {M.~D.}\ \bibnamefont {Johannes}}, \bibinfo
  {author} {\bibfnamefont {R.}~\bibnamefont {{Gonz{\'a}lez-Hern{\'a}ndez}}},\
  and\ \bibinfo {author} {\bibfnamefont {L.}~\bibnamefont {{\v S}mejkal}},\
  }\bibfield  {title} {\bibinfo {title} {Prediction of unconventional magnetism
  in doped {{FeSb}}$_2$},\ }\href {https://doi.org/10.1073/pnas.2108924118}
  {\bibfield  {journal} {\bibinfo  {journal} {Proc. Natl. Acad. Sci.}\ }\textbf
  {\bibinfo {volume} {118}},\ \bibinfo {pages} {e2108924118} (\bibinfo {year}
  {2021})}\BibitemShut {NoStop}%
\bibitem [{\citenamefont {Ma}\ \emph {et~al.}(2021)\citenamefont {Ma},
  \citenamefont {Hu}, \citenamefont {Li}, \citenamefont {Liu}, \citenamefont
  {Yao}, \citenamefont {Jia},\ and\ \citenamefont {Liu}}]{Ma2021:NC}%
  \BibitemOpen
  \bibfield  {author} {\bibinfo {author} {\bibfnamefont {H.-Y.}\ \bibnamefont
  {Ma}}, \bibinfo {author} {\bibfnamefont {M.}~\bibnamefont {Hu}}, \bibinfo
  {author} {\bibfnamefont {N.}~\bibnamefont {Li}}, \bibinfo {author}
  {\bibfnamefont {J.}~\bibnamefont {Liu}}, \bibinfo {author} {\bibfnamefont
  {W.}~\bibnamefont {Yao}}, \bibinfo {author} {\bibfnamefont {J.-F.}\
  \bibnamefont {Jia}},\ and\ \bibinfo {author} {\bibfnamefont {J.}~\bibnamefont
  {Liu}},\ }\bibfield  {title} {\bibinfo {title} {{Multifunctional
  antiferromagnetic materials with giant piezomagnetism and noncollinear spin
  current}},\ }\href {https://doi.org/10.1038/s41467-021-23127-7} {\bibfield
  {journal} {\bibinfo  {journal} {Nat. Commun.}\ }\textbf {\bibinfo {volume}
  {12}},\ \bibinfo {pages} {2846} (\bibinfo {year} {2021})}\BibitemShut
  {NoStop}%
\bibitem [{\citenamefont {Krempask{\'y}}\ \emph {et~al.}(2024)\citenamefont
  {Krempask{\'y}}, \citenamefont {{\v S}mejkal}, \citenamefont {D'Souza},
  \citenamefont {Hajlaoui}, \citenamefont {Springholz}, \citenamefont
  {Uhl{\'i}{\v r}ov{\'a}}, \citenamefont {Alarab}, \citenamefont
  {Constantinou}, \citenamefont {Strocov}, \citenamefont {Usanov},
  \citenamefont {Pudelko}, \citenamefont {{Gonz{\'a}lez-Hern{\'a}ndez}},
  \citenamefont {Birk~Hellenes}, \citenamefont {Jansa}, \citenamefont
  {Reichlov{\'a}}, \citenamefont {{\v S}ob{\'a}{\v n}}, \citenamefont
  {Gonzalez~Betancourt}, \citenamefont {Wadley}, \citenamefont {Sinova},
  \citenamefont {Kriegner}, \citenamefont {Min{\'a}r}, \citenamefont {Dil},\
  and\ \citenamefont {Jungwirth}}]{Krempasky2024:Nature}%
  \BibitemOpen
  \bibfield  {author} {\bibinfo {author} {\bibfnamefont {J.}~\bibnamefont
  {Krempask{\'y}}}, \bibinfo {author} {\bibfnamefont {L.}~\bibnamefont {{\v
  S}mejkal}}, \bibinfo {author} {\bibfnamefont {S.~W.}\ \bibnamefont
  {D'Souza}}, \bibinfo {author} {\bibfnamefont {M.}~\bibnamefont {Hajlaoui}},
  \bibinfo {author} {\bibfnamefont {G.}~\bibnamefont {Springholz}}, \bibinfo
  {author} {\bibfnamefont {K.}~\bibnamefont {Uhl{\'i}{\v r}ov{\'a}}}, \bibinfo
  {author} {\bibfnamefont {F.}~\bibnamefont {Alarab}}, \bibinfo {author}
  {\bibfnamefont {P.~C.}\ \bibnamefont {Constantinou}}, \bibinfo {author}
  {\bibfnamefont {V.}~\bibnamefont {Strocov}}, \bibinfo {author} {\bibfnamefont
  {D.}~\bibnamefont {Usanov}}, \bibinfo {author} {\bibfnamefont {W.~R.}\
  \bibnamefont {Pudelko}}, \bibinfo {author} {\bibfnamefont {R.}~\bibnamefont
  {{Gonz{\'a}lez-Hern{\'a}ndez}}}, \bibinfo {author} {\bibfnamefont
  {A.}~\bibnamefont {Birk~Hellenes}}, \bibinfo {author} {\bibfnamefont
  {Z.}~\bibnamefont {Jansa}}, \bibinfo {author} {\bibfnamefont
  {H.}~\bibnamefont {Reichlov{\'a}}}, \bibinfo {author} {\bibfnamefont
  {Z.}~\bibnamefont {{\v S}ob{\'a}{\v n}}}, \bibinfo {author} {\bibfnamefont
  {R.~D.}\ \bibnamefont {Gonzalez~Betancourt}}, \bibinfo {author}
  {\bibfnamefont {P.}~\bibnamefont {Wadley}}, \bibinfo {author} {\bibfnamefont
  {J.}~\bibnamefont {Sinova}}, \bibinfo {author} {\bibfnamefont
  {D.}~\bibnamefont {Kriegner}}, \bibinfo {author} {\bibfnamefont
  {J.}~\bibnamefont {Min{\'a}r}}, \bibinfo {author} {\bibfnamefont {J.~H.}\
  \bibnamefont {Dil}},\ and\ \bibinfo {author} {\bibfnamefont {T.}~\bibnamefont
  {Jungwirth}},\ }\bibfield  {title} {\bibinfo {title} {Altermagnetic lifting
  of {{Kramers}} spin degeneracy},\ }\href
  {https://doi.org/10.1038/s41586-023-06907-7} {\bibfield  {journal} {\bibinfo
  {journal} {Nature}\ }\textbf {\bibinfo {volume} {626}},\ \bibinfo {pages}
  {517} (\bibinfo {year} {2024})}\BibitemShut {NoStop}%
\bibitem [{\citenamefont {Zhou}\ \emph {et~al.}(2025)\citenamefont {Zhou},
  \citenamefont {Cheng}, \citenamefont {Hu}, \citenamefont {Chu}, \citenamefont
  {Bai}, \citenamefont {Han}, \citenamefont {Liu}, \citenamefont {Pan},\ and\
  \citenamefont {Song}}]{Zhou2025:Nature}%
  \BibitemOpen
  \bibfield  {author} {\bibinfo {author} {\bibfnamefont {Z.}~\bibnamefont
  {Zhou}}, \bibinfo {author} {\bibfnamefont {X.}~\bibnamefont {Cheng}},
  \bibinfo {author} {\bibfnamefont {M.}~\bibnamefont {Hu}}, \bibinfo {author}
  {\bibfnamefont {R.}~\bibnamefont {Chu}}, \bibinfo {author} {\bibfnamefont
  {H.}~\bibnamefont {Bai}}, \bibinfo {author} {\bibfnamefont {L.}~\bibnamefont
  {Han}}, \bibinfo {author} {\bibfnamefont {J.}~\bibnamefont {Liu}}, \bibinfo
  {author} {\bibfnamefont {F.}~\bibnamefont {Pan}},\ and\ \bibinfo {author}
  {\bibfnamefont {C.}~\bibnamefont {Song}},\ }\bibfield  {title} {\bibinfo
  {title} {Manipulation of the altermagnetic order in {{CrSb}} via crystal
  symmetry},\ }\href {https://doi.org/10.1038/s41586-024-08436-3} {\bibfield
  {journal} {\bibinfo  {journal} {Nature}\ }\textbf {\bibinfo {volume} {638}},\
  \bibinfo {pages} {645} (\bibinfo {year} {2025})}\BibitemShut {NoStop}%
\bibitem [{\citenamefont {Zhang}\ \emph
  {et~al.}(2025{\natexlab{a}})\citenamefont {Zhang}, \citenamefont {Cheng},
  \citenamefont {Yin}, \citenamefont {Liu}, \citenamefont {Deng}, \citenamefont
  {Qiao}, \citenamefont {Shi}, \citenamefont {Zhang}, \citenamefont {Lin},
  \citenamefont {Liu}, \citenamefont {Ye}, \citenamefont {Huang}, \citenamefont
  {Meng}, \citenamefont {Zhang}, \citenamefont {Okuda}, \citenamefont
  {Shimada}, \citenamefont {Cui}, \citenamefont {Zhao}, \citenamefont {Cao},
  \citenamefont {Qiao}, \citenamefont {Liu},\ and\ \citenamefont
  {Chen}}]{Zhang2025:NP}%
  \BibitemOpen
  \bibfield  {author} {\bibinfo {author} {\bibfnamefont {F.}~\bibnamefont
  {Zhang}}, \bibinfo {author} {\bibfnamefont {X.}~\bibnamefont {Cheng}},
  \bibinfo {author} {\bibfnamefont {Z.}~\bibnamefont {Yin}}, \bibinfo {author}
  {\bibfnamefont {C.}~\bibnamefont {Liu}}, \bibinfo {author} {\bibfnamefont
  {L.}~\bibnamefont {Deng}}, \bibinfo {author} {\bibfnamefont {Y.}~\bibnamefont
  {Qiao}}, \bibinfo {author} {\bibfnamefont {Z.}~\bibnamefont {Shi}}, \bibinfo
  {author} {\bibfnamefont {S.}~\bibnamefont {Zhang}}, \bibinfo {author}
  {\bibfnamefont {J.}~\bibnamefont {Lin}}, \bibinfo {author} {\bibfnamefont
  {Z.}~\bibnamefont {Liu}}, \bibinfo {author} {\bibfnamefont {M.}~\bibnamefont
  {Ye}}, \bibinfo {author} {\bibfnamefont {Y.}~\bibnamefont {Huang}}, \bibinfo
  {author} {\bibfnamefont {X.}~\bibnamefont {Meng}}, \bibinfo {author}
  {\bibfnamefont {C.}~\bibnamefont {Zhang}}, \bibinfo {author} {\bibfnamefont
  {T.}~\bibnamefont {Okuda}}, \bibinfo {author} {\bibfnamefont
  {K.}~\bibnamefont {Shimada}}, \bibinfo {author} {\bibfnamefont
  {S.}~\bibnamefont {Cui}}, \bibinfo {author} {\bibfnamefont {Y.}~\bibnamefont
  {Zhao}}, \bibinfo {author} {\bibfnamefont {G.-H.}\ \bibnamefont {Cao}},
  \bibinfo {author} {\bibfnamefont {S.}~\bibnamefont {Qiao}}, \bibinfo {author}
  {\bibfnamefont {J.}~\bibnamefont {Liu}},\ and\ \bibinfo {author}
  {\bibfnamefont {C.}~\bibnamefont {Chen}},\ }\bibfield  {title} {\bibinfo
  {title} {Crystal-symmetry-paired spin--valley locking in a layered
  room-temperature metallic altermagnet candidate},\ }\href
  {https://doi.org/10.1038/s41567-025-02864-2} {\bibfield  {journal} {\bibinfo
  {journal} {Nat. Phys.}\ }\textbf {\bibinfo {volume} {21}},\ \bibinfo {pages}
  {760} (\bibinfo {year} {2025}{\natexlab{a}})}\BibitemShut {NoStop}%
\bibitem [{\citenamefont {Jiang}\ \emph {et~al.}(2025)\citenamefont {Jiang},
  \citenamefont {Hu}, \citenamefont {Bai}, \citenamefont {Song}, \citenamefont
  {Mu}, \citenamefont {Qu}, \citenamefont {Li}, \citenamefont {Zhu},
  \citenamefont {Pi}, \citenamefont {Wei}, \citenamefont {Sun}, \citenamefont
  {Huang}, \citenamefont {Zheng}, \citenamefont {Peng}, \citenamefont {He},
  \citenamefont {Li}, \citenamefont {Luo}, \citenamefont {Li}, \citenamefont
  {Chen}, \citenamefont {Li}, \citenamefont {Weng},\ and\ \citenamefont
  {Qian}}]{Jiang2025:NP}%
  \BibitemOpen
  \bibfield  {author} {\bibinfo {author} {\bibfnamefont {B.}~\bibnamefont
  {Jiang}}, \bibinfo {author} {\bibfnamefont {M.}~\bibnamefont {Hu}}, \bibinfo
  {author} {\bibfnamefont {J.}~\bibnamefont {Bai}}, \bibinfo {author}
  {\bibfnamefont {Z.}~\bibnamefont {Song}}, \bibinfo {author} {\bibfnamefont
  {C.}~\bibnamefont {Mu}}, \bibinfo {author} {\bibfnamefont {G.}~\bibnamefont
  {Qu}}, \bibinfo {author} {\bibfnamefont {W.}~\bibnamefont {Li}}, \bibinfo
  {author} {\bibfnamefont {W.}~\bibnamefont {Zhu}}, \bibinfo {author}
  {\bibfnamefont {H.}~\bibnamefont {Pi}}, \bibinfo {author} {\bibfnamefont
  {Z.}~\bibnamefont {Wei}}, \bibinfo {author} {\bibfnamefont {Y.-J.}\
  \bibnamefont {Sun}}, \bibinfo {author} {\bibfnamefont {Y.}~\bibnamefont
  {Huang}}, \bibinfo {author} {\bibfnamefont {X.}~\bibnamefont {Zheng}},
  \bibinfo {author} {\bibfnamefont {Y.}~\bibnamefont {Peng}}, \bibinfo {author}
  {\bibfnamefont {L.}~\bibnamefont {He}}, \bibinfo {author} {\bibfnamefont
  {S.}~\bibnamefont {Li}}, \bibinfo {author} {\bibfnamefont {J.}~\bibnamefont
  {Luo}}, \bibinfo {author} {\bibfnamefont {Z.}~\bibnamefont {Li}}, \bibinfo
  {author} {\bibfnamefont {G.}~\bibnamefont {Chen}}, \bibinfo {author}
  {\bibfnamefont {H.}~\bibnamefont {Li}}, \bibinfo {author} {\bibfnamefont
  {H.}~\bibnamefont {Weng}},\ and\ \bibinfo {author} {\bibfnamefont
  {T.}~\bibnamefont {Qian}},\ }\bibfield  {title} {\bibinfo {title} {A metallic
  room-temperature \$d\$-wave altermagnet},\ }\href
  {https://doi.org/10.1038/s41567-025-02822-y} {\bibfield  {journal} {\bibinfo
  {journal} {Nat. Phys.}\ }\textbf {\bibinfo {volume} {21}},\ \bibinfo {pages}
  {754} (\bibinfo {year} {2025})}\BibitemShut {NoStop}%
\bibitem [{\citenamefont {Wang}\ \emph {et~al.}(2025)\citenamefont {Wang},
  \citenamefont {Yu}, \citenamefont {Cheng}, \citenamefont {Xiao},
  \citenamefont {Ma}, \citenamefont {Quan}, \citenamefont {Song}, \citenamefont
  {Zhang}, \citenamefont {Zhang}, \citenamefont {Ma}, \citenamefont {Liu},
  \citenamefont {Yadav}, \citenamefont {Shi}, \citenamefont {Wang},
  \citenamefont {Niu}, \citenamefont {Gao}, \citenamefont {Xiang},
  \citenamefont {Liu}, \citenamefont {Wang},\ and\ \citenamefont
  {Chen}}]{Wang2025:arXiv}%
  \BibitemOpen
  \bibfield  {author} {\bibinfo {author} {\bibfnamefont {Z.}~\bibnamefont
  {Wang}}, \bibinfo {author} {\bibfnamefont {S.}~\bibnamefont {Yu}}, \bibinfo
  {author} {\bibfnamefont {X.}~\bibnamefont {Cheng}}, \bibinfo {author}
  {\bibfnamefont {X.}~\bibnamefont {Xiao}}, \bibinfo {author} {\bibfnamefont
  {W.}~\bibnamefont {Ma}}, \bibinfo {author} {\bibfnamefont {F.}~\bibnamefont
  {Quan}}, \bibinfo {author} {\bibfnamefont {H.}~\bibnamefont {Song}}, \bibinfo
  {author} {\bibfnamefont {K.}~\bibnamefont {Zhang}}, \bibinfo {author}
  {\bibfnamefont {Y.}~\bibnamefont {Zhang}}, \bibinfo {author} {\bibfnamefont
  {Y.}~\bibnamefont {Ma}}, \bibinfo {author} {\bibfnamefont {W.}~\bibnamefont
  {Liu}}, \bibinfo {author} {\bibfnamefont {P.}~\bibnamefont {Yadav}}, \bibinfo
  {author} {\bibfnamefont {X.}~\bibnamefont {Shi}}, \bibinfo {author}
  {\bibfnamefont {Z.}~\bibnamefont {Wang}}, \bibinfo {author} {\bibfnamefont
  {Q.}~\bibnamefont {Niu}}, \bibinfo {author} {\bibfnamefont {Y.}~\bibnamefont
  {Gao}}, \bibinfo {author} {\bibfnamefont {B.}~\bibnamefont {Xiang}}, \bibinfo
  {author} {\bibfnamefont {J.}~\bibnamefont {Liu}}, \bibinfo {author}
  {\bibfnamefont {Z.}~\bibnamefont {Wang}},\ and\ \bibinfo {author}
  {\bibfnamefont {X.}~\bibnamefont {Chen}},\ }\bibfield  {title} {\bibinfo
  {title} {Atomic-scale spin sensing of a {2D} $d$-wave altermagnet via helical
  tunneling},\ }\href {https://arxiv.org/abs/2512.23290} {\bibfield  {journal}
  {\bibinfo  {journal} {arXiv:2512.23290}\ } (\bibinfo {year}
  {2025})}\BibitemShut {NoStop}%
\bibitem [{\citenamefont {Fu}\ \emph {et~al.}(2025{\natexlab{a}})\citenamefont
  {Fu}, \citenamefont {Yang}, \citenamefont {Xiao}, \citenamefont {Wang},
  \citenamefont {Wang}, \citenamefont {Yao}, \citenamefont {Xue},\ and\
  \citenamefont {Li}}]{FuD2025:arXiv}%
  \BibitemOpen
  \bibfield  {author} {\bibinfo {author} {\bibfnamefont {D.}~\bibnamefont
  {Fu}}, \bibinfo {author} {\bibfnamefont {L.}~\bibnamefont {Yang}}, \bibinfo
  {author} {\bibfnamefont {K.}~\bibnamefont {Xiao}}, \bibinfo {author}
  {\bibfnamefont {Y.}~\bibnamefont {Wang}}, \bibinfo {author} {\bibfnamefont
  {Z.}~\bibnamefont {Wang}}, \bibinfo {author} {\bibfnamefont {Y.}~\bibnamefont
  {Yao}}, \bibinfo {author} {\bibfnamefont {Q.-K.}\ \bibnamefont {Xue}},\ and\
  \bibinfo {author} {\bibfnamefont {W.}~\bibnamefont {Li}},\ }\bibfield
  {title} {\bibinfo {title} {Atomic-scale visualization of d-wave
  altermagnetism},\ }\href {https://arxiv.org/abs/2512.24114} {\bibfield
  {journal} {\bibinfo  {journal} {arXiv.2512.24114}\ } (\bibinfo {year}
  {2025}{\natexlab{a}})}\BibitemShut {NoStop}%
\bibitem [{\citenamefont {{\v S}mejkal}\ \emph {et~al.}(2022)\citenamefont {{\v
  S}mejkal}, \citenamefont {Sinova},\ and\ \citenamefont
  {Jungwirth}}]{Smejkal2022:PRX}%
  \BibitemOpen
  \bibfield  {author} {\bibinfo {author} {\bibfnamefont {L.}~\bibnamefont {{\v
  S}mejkal}}, \bibinfo {author} {\bibfnamefont {J.}~\bibnamefont {Sinova}},\
  and\ \bibinfo {author} {\bibfnamefont {T.}~\bibnamefont {Jungwirth}},\
  }\bibfield  {title} {\bibinfo {title} {Beyond {{Conventional Ferromagnetism}}
  and {{Antiferromagnetism}}: {{A Phase}} with {{Nonrelativistic Spin}} and
  {{Crystal Rotation Symmetry}}},\ }\href
  {https://doi.org/10.1103/PhysRevX.12.031042} {\bibfield  {journal} {\bibinfo
  {journal} {Phys. Rev. X}\ }\textbf {\bibinfo {volume} {12}},\ \bibinfo
  {pages} {031042} (\bibinfo {year} {2022})}\BibitemShut {NoStop}%
\bibitem [{\citenamefont {{\v{S}}mejkal}\ \emph {et~al.}(2022)\citenamefont
  {{\v{S}}mejkal}, \citenamefont {Sinova},\ and\ \citenamefont
  {Jungwirth}}]{Smejkal2022:PRX2}%
  \BibitemOpen
  \bibfield  {author} {\bibinfo {author} {\bibfnamefont {L.}~\bibnamefont
  {{\v{S}}mejkal}}, \bibinfo {author} {\bibfnamefont {J.}~\bibnamefont
  {Sinova}},\ and\ \bibinfo {author} {\bibfnamefont {T.}~\bibnamefont
  {Jungwirth}},\ }\bibfield  {title} {\bibinfo {title} {Emerging {{Research
  Landscape}} of {{Altermagnetism}}},\ }\href
  {https://doi.org/10.1103/PhysRevX.12.040501} {\bibfield  {journal} {\bibinfo
  {journal} {Phys. Rev. X}\ }\textbf {\bibinfo {volume} {12}},\ \bibinfo
  {pages} {040501} (\bibinfo {year} {2022})}\BibitemShut {NoStop}%
\bibitem [{\citenamefont {Bai}\ \emph {et~al.}(2024)\citenamefont {Bai},
  \citenamefont {Feng}, \citenamefont {Liu}, \citenamefont {{\v{S}}mejkal},
  \citenamefont {Mokrousov},\ and\ \citenamefont {Yao}}]{Bai2024:AFM}%
  \BibitemOpen
  \bibfield  {author} {\bibinfo {author} {\bibfnamefont {L.}~\bibnamefont
  {Bai}}, \bibinfo {author} {\bibfnamefont {W.}~\bibnamefont {Feng}}, \bibinfo
  {author} {\bibfnamefont {S.}~\bibnamefont {Liu}}, \bibinfo {author}
  {\bibfnamefont {L.}~\bibnamefont {{\v{S}}mejkal}}, \bibinfo {author}
  {\bibfnamefont {Y.}~\bibnamefont {Mokrousov}},\ and\ \bibinfo {author}
  {\bibfnamefont {Y.}~\bibnamefont {Yao}},\ }\bibfield  {title} {\bibinfo
  {title} {Altermagnetism: {{Exploring}} new frontiers in magnetism and
  spintronics},\ }\href {https://doi.org/10.1002/adfm.202409327} {\bibfield
  {journal} {\bibinfo  {journal} {Adv. Funct. Mater.}\ }\textbf {\bibinfo
  {volume} {34}},\ \bibinfo {pages} {2409327} (\bibinfo {year}
  {2024})}\BibitemShut {NoStop}%
\bibitem [{\citenamefont {Song}\ \emph {et~al.}(2025)\citenamefont {Song},
  \citenamefont {Bai}, \citenamefont {Zhou}, \citenamefont {Han}, \citenamefont
  {Reichlova}, \citenamefont {Dil}, \citenamefont {Liu}, \citenamefont {Chen},\
  and\ \citenamefont {Pan}}]{Song2025:NRM}%
  \BibitemOpen
  \bibfield  {author} {\bibinfo {author} {\bibfnamefont {C.}~\bibnamefont
  {Song}}, \bibinfo {author} {\bibfnamefont {H.}~\bibnamefont {Bai}}, \bibinfo
  {author} {\bibfnamefont {Z.}~\bibnamefont {Zhou}}, \bibinfo {author}
  {\bibfnamefont {L.}~\bibnamefont {Han}}, \bibinfo {author} {\bibfnamefont
  {H.}~\bibnamefont {Reichlova}}, \bibinfo {author} {\bibfnamefont {J.~H.}\
  \bibnamefont {Dil}}, \bibinfo {author} {\bibfnamefont {J.}~\bibnamefont
  {Liu}}, \bibinfo {author} {\bibfnamefont {X.}~\bibnamefont {Chen}},\ and\
  \bibinfo {author} {\bibfnamefont {F.}~\bibnamefont {Pan}},\ }\bibfield
  {title} {\bibinfo {title} {Altermagnets as a new class of functional
  materials},\ }\href {https://doi.org/10.1038/s41578-025-00779-1} {\bibfield
  {journal} {\bibinfo  {journal} {Nat. Rev. Mater.}\ }\textbf {\bibinfo
  {volume} {10}},\ \bibinfo {pages} {473} (\bibinfo {year} {2025})}\BibitemShut
  {NoStop}%
\bibitem [{\citenamefont {Bhowal}\ and\ \citenamefont
  {Bose}(2025)}]{Bhowal2025:arXiv}%
  \BibitemOpen
  \bibfield  {author} {\bibinfo {author} {\bibfnamefont {S.}~\bibnamefont
  {Bhowal}}\ and\ \bibinfo {author} {\bibfnamefont {A.}~\bibnamefont {Bose}},\
  }\bibfield  {title} {\bibinfo {title} {Non-relativistic spin splitting:
  {Features} and functionalities},\ }\href {https://arxiv.org/abs/2510.20306}
  {\bibfield  {journal} {\bibinfo  {journal} {arXiv: 2510.20306}\ } (\bibinfo
  {year} {2025})}\BibitemShut {NoStop}%
\bibitem [{\citenamefont {Cheong}\ and\ \citenamefont
  {Huang}(2025)}]{Cheong2025:NQM}%
  \BibitemOpen
  \bibfield  {author} {\bibinfo {author} {\bibfnamefont {S.-W.}\ \bibnamefont
  {Cheong}}\ and\ \bibinfo {author} {\bibfnamefont {F.-T.}\ \bibnamefont
  {Huang}},\ }\bibfield  {title} {\bibinfo {title} {Altermagnetism
  classification},\ }\href {https://doi.org/10.1038/s41535-025-00756-5}
  {\bibfield  {journal} {\bibinfo  {journal} {npj Quantum Mater.}\ }\textbf
  {\bibinfo {volume} {10}},\ \bibinfo {pages} {38} (\bibinfo {year}
  {2025})}\BibitemShut {NoStop}%
\bibitem [{\citenamefont {Guo}\ \emph {et~al.}(2025)\citenamefont {Guo},
  \citenamefont {Wang}, \citenamefont {Wang}, \citenamefont {Zhang},
  \citenamefont {Zhou},\ and\ \citenamefont {Cheng}}]{Guo2025:AM}%
  \BibitemOpen
  \bibfield  {author} {\bibinfo {author} {\bibfnamefont {Z.}~\bibnamefont
  {Guo}}, \bibinfo {author} {\bibfnamefont {X.}~\bibnamefont {Wang}}, \bibinfo
  {author} {\bibfnamefont {W.}~\bibnamefont {Wang}}, \bibinfo {author}
  {\bibfnamefont {G.}~\bibnamefont {Zhang}}, \bibinfo {author} {\bibfnamefont
  {X.}~\bibnamefont {Zhou}},\ and\ \bibinfo {author} {\bibfnamefont
  {Z.}~\bibnamefont {Cheng}},\ }\bibfield  {title} {\bibinfo {title}
  {Spin-polarized antiferromagnets for spintronics},\ }\href
  {https://doi.org/10.1002/adma.202505779} {\bibfield  {journal} {\bibinfo
  {journal} {Adv. Mater.}\ }\textbf {\bibinfo {volume} {37}},\ \bibinfo {pages}
  {2505779} (\bibinfo {year} {2025})}\BibitemShut {NoStop}%
\bibitem [{\citenamefont {Yu}\ \emph {et~al.}(2025)\citenamefont {Yu},
  \citenamefont {Ji}, \citenamefont {Luo}, \citenamefont {Gong},\ and\
  \citenamefont {Xiang}}]{Yu2025:AM}%
  \BibitemOpen
  \bibfield  {author} {\bibinfo {author} {\bibfnamefont {H.}~\bibnamefont
  {Yu}}, \bibinfo {author} {\bibfnamefont {J.}~\bibnamefont {Ji}}, \bibinfo
  {author} {\bibfnamefont {W.}~\bibnamefont {Luo}}, \bibinfo {author}
  {\bibfnamefont {X.}~\bibnamefont {Gong}},\ and\ \bibinfo {author}
  {\bibfnamefont {H.}~\bibnamefont {Xiang}},\ }\bibfield  {title} {\bibinfo
  {title} {Recent advances in unconventional ferroelectrics and
  multiferroics},\ }\href {https://doi.org/10.1002/adma.202507070} {\bibfield
  {journal} {\bibinfo  {journal} {Adv. Mater.}\ }\textbf {\bibinfo {volume}
  {n/a}},\ \bibinfo {pages} {e07070} (\bibinfo {year} {2025})}\BibitemShut
  {NoStop}%
\bibitem [{\citenamefont {Fukaya}\ \emph {et~al.}(2025)\citenamefont {Fukaya},
  \citenamefont {Lu}, \citenamefont {Yada}, \citenamefont {Tanaka},\ and\
  \citenamefont {Cayao}}]{Fukaya2025:JPCM}%
  \BibitemOpen
  \bibfield  {author} {\bibinfo {author} {\bibfnamefont {Y.}~\bibnamefont
  {Fukaya}}, \bibinfo {author} {\bibfnamefont {B.}~\bibnamefont {Lu}}, \bibinfo
  {author} {\bibfnamefont {K.}~\bibnamefont {Yada}}, \bibinfo {author}
  {\bibfnamefont {Y.}~\bibnamefont {Tanaka}},\ and\ \bibinfo {author}
  {\bibfnamefont {J.}~\bibnamefont {Cayao}},\ }\bibfield  {title} {\bibinfo
  {title} {Superconducting phenomena in systems with unconventional magnets},\
  }\href {https://doi.org/10.1088/1361-648X/adf1cf} {\bibfield  {journal}
  {\bibinfo  {journal} {J. Phys.: Condens. Matter}\ }\textbf {\bibinfo {volume}
  {37}},\ \bibinfo {pages} {313003} (\bibinfo {year} {2025})}\BibitemShut
  {NoStop}%
\bibitem [{\citenamefont {Duan}\ \emph {et~al.}(2025)\citenamefont {Duan},
  \citenamefont {Zhang}, \citenamefont {Zhu}, \citenamefont {Liu},
  \citenamefont {Zhang}, \citenamefont {\ifmmode \check{Z}\else
  \v{Z}\fi{}uti\ifmmode~\acute{c}\else \'{c}\fi{}},\ and\ \citenamefont
  {Zhou}}]{Duan2025:PRL}%
  \BibitemOpen
  \bibfield  {author} {\bibinfo {author} {\bibfnamefont {X.}~\bibnamefont
  {Duan}}, \bibinfo {author} {\bibfnamefont {J.}~\bibnamefont {Zhang}},
  \bibinfo {author} {\bibfnamefont {Z.}~\bibnamefont {Zhu}}, \bibinfo {author}
  {\bibfnamefont {Y.}~\bibnamefont {Liu}}, \bibinfo {author} {\bibfnamefont
  {Z.}~\bibnamefont {Zhang}}, \bibinfo {author} {\bibfnamefont
  {I.}~\bibnamefont {\ifmmode \check{Z}\else
  \v{Z}\fi{}uti\ifmmode~\acute{c}\else \'{c}\fi{}}},\ and\ \bibinfo {author}
  {\bibfnamefont {T.}~\bibnamefont {Zhou}},\ }\bibfield  {title} {\bibinfo
  {title} {Antiferroelectric {{Altermagnets}}: {{Antiferroelectricity Alters
  Magnets}}},\ }\href {https://doi.org/10.1103/PhysRevLett.134.106801}
  {\bibfield  {journal} {\bibinfo  {journal} {Phys. Rev. Lett.}\ }\textbf
  {\bibinfo {volume} {134}},\ \bibinfo {pages} {106801} (\bibinfo {year}
  {2025})}\BibitemShut {NoStop}%
\bibitem [{\citenamefont {Zhu}\ \emph {et~al.}(2025{\natexlab{a}})\citenamefont
  {Zhu}, \citenamefont {Duan}, \citenamefont {Zhang}, \citenamefont {Hao},
  \citenamefont {{\v Z}uti{\'c}},\ and\ \citenamefont {Zhou}}]{Zhu2025:NL}%
  \BibitemOpen
  \bibfield  {author} {\bibinfo {author} {\bibfnamefont {Z.}~\bibnamefont
  {Zhu}}, \bibinfo {author} {\bibfnamefont {X.}~\bibnamefont {Duan}}, \bibinfo
  {author} {\bibfnamefont {J.}~\bibnamefont {Zhang}}, \bibinfo {author}
  {\bibfnamefont {B.}~\bibnamefont {Hao}}, \bibinfo {author} {\bibfnamefont
  {I.}~\bibnamefont {{\v Z}uti{\'c}}},\ and\ \bibinfo {author} {\bibfnamefont
  {T.}~\bibnamefont {Zhou}},\ }\bibfield  {title} {\bibinfo {title}
  {Two-dimensional ferroelectric altermagnets: {{From}} model to material
  realization},\ }\href {10.1021/acs.nanolett.5c02121} {\bibfield  {journal}
  {\bibinfo  {journal} {Nano Lett.}\ }\textbf {\bibinfo {volume} {25}},\
  \bibinfo {pages} {9456} (\bibinfo {year} {2025}{\natexlab{a}})}\BibitemShut
  {NoStop}%
\bibitem [{\citenamefont {Gu}\ \emph {et~al.}(2025)\citenamefont {Gu},
  \citenamefont {Liu}, \citenamefont {Zhu}, \citenamefont {Yananose},
  \citenamefont {Chen}, \citenamefont {Hu}, \citenamefont {Stroppa},\ and\
  \citenamefont {Liu}}]{Gu2025:PRL}%
  \BibitemOpen
  \bibfield  {author} {\bibinfo {author} {\bibfnamefont {M.}~\bibnamefont
  {Gu}}, \bibinfo {author} {\bibfnamefont {Y.}~\bibnamefont {Liu}}, \bibinfo
  {author} {\bibfnamefont {H.}~\bibnamefont {Zhu}}, \bibinfo {author}
  {\bibfnamefont {K.}~\bibnamefont {Yananose}}, \bibinfo {author}
  {\bibfnamefont {X.}~\bibnamefont {Chen}}, \bibinfo {author} {\bibfnamefont
  {Y.}~\bibnamefont {Hu}}, \bibinfo {author} {\bibfnamefont {A.}~\bibnamefont
  {Stroppa}},\ and\ \bibinfo {author} {\bibfnamefont {Q.}~\bibnamefont {Liu}},\
  }\bibfield  {title} {\bibinfo {title} {Ferroelectric {{Switchable
  Altermagnetism}}},\ }\href {https://doi.org/10.1103/PhysRevLett.134.106802}
  {\bibfield  {journal} {\bibinfo  {journal} {Phys. Rev. Lett.}\ }\textbf
  {\bibinfo {volume} {134}},\ \bibinfo {pages} {106802} (\bibinfo {year}
  {2025})}\BibitemShut {NoStop}%
\bibitem [{\citenamefont {Sun}\ \emph {et~al.}(2025)\citenamefont {Sun},
  \citenamefont {Yang}, \citenamefont {Wang}, \citenamefont {Liu},
  \citenamefont {Wang}, \citenamefont {Huang},\ and\ \citenamefont
  {Cheng}}]{Sun2025:AM}%
  \BibitemOpen
  \bibfield  {author} {\bibinfo {author} {\bibfnamefont {W.}~\bibnamefont
  {Sun}}, \bibinfo {author} {\bibfnamefont {C.}~\bibnamefont {Yang}}, \bibinfo
  {author} {\bibfnamefont {W.}~\bibnamefont {Wang}}, \bibinfo {author}
  {\bibfnamefont {Y.}~\bibnamefont {Liu}}, \bibinfo {author} {\bibfnamefont
  {X.}~\bibnamefont {Wang}}, \bibinfo {author} {\bibfnamefont {S.}~\bibnamefont
  {Huang}},\ and\ \bibinfo {author} {\bibfnamefont {Z.}~\bibnamefont {Cheng}},\
  }\bibfield  {title} {\bibinfo {title} {Proposing altermagnetic-ferroelectric
  type-{{III}} multiferroics with robust magnetoelectric coupling},\ }\href
  {https://doi.org/https://doi.org/10.1002/adma.202502575} {\bibfield
  {journal} {\bibinfo  {journal} {Adv. Mater.}\ }\textbf {\bibinfo {volume}
  {37}},\ \bibinfo {pages} {2502575} (\bibinfo {year} {2025})}\BibitemShut
  {NoStop}%
\bibitem [{\citenamefont {Liu}\ \emph {et~al.}(2025{\natexlab{a}})\citenamefont
  {Liu}, \citenamefont {Dai},\ and\ \citenamefont {Bl{\"u}gel}}]{Liu2025:NP}%
  \BibitemOpen
  \bibfield  {author} {\bibinfo {author} {\bibfnamefont {Q.}~\bibnamefont
  {Liu}}, \bibinfo {author} {\bibfnamefont {X.}~\bibnamefont {Dai}},\ and\
  \bibinfo {author} {\bibfnamefont {S.}~\bibnamefont {Bl{\"u}gel}},\ }\bibfield
   {title} {\bibinfo {title} {Different facets of unconventional magnetism},\
  }\href {https://doi.org/10.1038/s41567-024-02750-3} {\bibfield  {journal}
  {\bibinfo  {journal} {Nat. Phys.}\ }\textbf {\bibinfo {volume} {21}},\
  \bibinfo {pages} {329} (\bibinfo {year} {2025}{\natexlab{a}})}\BibitemShut
  {NoStop}%
\bibitem [{\citenamefont {Jungwirth}\ \emph {et~al.}(2026)\citenamefont
  {Jungwirth}, \citenamefont {Sinova}, \citenamefont {Fernandes}, \citenamefont
  {Liu}, \citenamefont {Watanabe}, \citenamefont {Murakami}, \citenamefont
  {Nakatsuji},\ and\ \citenamefont {{\v{S}}mejkal}}]{Jungwirth2026:Nature}%
  \BibitemOpen
  \bibfield  {author} {\bibinfo {author} {\bibfnamefont {T.}~\bibnamefont
  {Jungwirth}}, \bibinfo {author} {\bibfnamefont {J.}~\bibnamefont {Sinova}},
  \bibinfo {author} {\bibfnamefont {R.~M.}\ \bibnamefont {Fernandes}}, \bibinfo
  {author} {\bibfnamefont {Q.}~\bibnamefont {Liu}}, \bibinfo {author}
  {\bibfnamefont {H.}~\bibnamefont {Watanabe}}, \bibinfo {author}
  {\bibfnamefont {S.}~\bibnamefont {Murakami}}, \bibinfo {author}
  {\bibfnamefont {S.}~\bibnamefont {Nakatsuji}},\ and\ \bibinfo {author}
  {\bibfnamefont {L.}~\bibnamefont {{\v{S}}mejkal}},\ }\bibfield  {title}
  {\bibinfo {title} {Symmetry, microscopy and spectroscopy signatures of
  altermagnetism},\ }\href
  {https://doi.org/https://doi.org/10.1038/s41586-025-09883-2} {\bibfield
  {journal} {\bibinfo  {journal} {Nature}\ }\textbf {\bibinfo {volume} {649}},\
  \bibinfo {pages} {837} (\bibinfo {year} {2026})}\BibitemShut {NoStop}%
\bibitem [{\citenamefont {Mazin}\ \emph {et~al.}(2023)\citenamefont {Mazin},
  \citenamefont {Gonz{\'a}lez-Hern{\'a}ndez},\ and\ \citenamefont
  {{\v{S}}mejkal}}]{Mazin2023:arXiv}%
  \BibitemOpen
  \bibfield  {author} {\bibinfo {author} {\bibfnamefont {I.}~\bibnamefont
  {Mazin}}, \bibinfo {author} {\bibfnamefont {R.}~\bibnamefont
  {Gonz{\'a}lez-Hern{\'a}ndez}},\ and\ \bibinfo {author} {\bibfnamefont
  {L.}~\bibnamefont {{\v{S}}mejkal}},\ }\bibfield  {title} {\bibinfo {title}
  {Induced monolayer altermagnetism in {MnP(S,Se)}$_3$ and {FeSe}},\ }\href
  {https://doi.org/10.48550/arXiv.2309.02355} {\bibfield  {journal} {\bibinfo
  {journal} {arXiv:2309.02355}\ } (\bibinfo {year} {2023})}\BibitemShut
  {NoStop}%
\bibitem [{\citenamefont {Guo}\ \emph {et~al.}(2023)\citenamefont {Guo},
  \citenamefont {Liu},\ and\ \citenamefont {Lu}}]{Guo2023:NCM}%
  \BibitemOpen
  \bibfield  {author} {\bibinfo {author} {\bibfnamefont {P.-J.}\ \bibnamefont
  {Guo}}, \bibinfo {author} {\bibfnamefont {Z.-X.}\ \bibnamefont {Liu}},\ and\
  \bibinfo {author} {\bibfnamefont {Z.-Y.}\ \bibnamefont {Lu}},\ }\bibfield
  {title} {\bibinfo {title} {Quantum anomalous {Hall} effect in collinear
  antiferromagnetism},\ }\href {https://doi.org/10.1038/s41524-023-01025-4}
  {\bibfield  {journal} {\bibinfo  {journal} {npj Comput. Mater.}\ }\textbf
  {\bibinfo {volume} {9}},\ \bibinfo {pages} {70} (\bibinfo {year}
  {2023})}\BibitemShut {NoStop}%
\bibitem [{\citenamefont {Chen}\ \emph {et~al.}(2023)\citenamefont {Chen},
  \citenamefont {Wang}, \citenamefont {Li},\ and\ \citenamefont
  {Sanyal}}]{Chen2023:APL}%
  \BibitemOpen
  \bibfield  {author} {\bibinfo {author} {\bibfnamefont {X.}~\bibnamefont
  {Chen}}, \bibinfo {author} {\bibfnamefont {D.}~\bibnamefont {Wang}}, \bibinfo
  {author} {\bibfnamefont {L.}~\bibnamefont {Li}},\ and\ \bibinfo {author}
  {\bibfnamefont {B.}~\bibnamefont {Sanyal}},\ }\bibfield  {title} {\bibinfo
  {title} {{Giant spin-splitting and tunable spin-momentum locked transport in
  room temperature collinear antiferromagnetic semimetallic CrO monolayer}},\
  }\href {https://doi.org/10.1063/5.0147450} {\bibfield  {journal} {\bibinfo
  {journal} {Appl. Phys. Lett.}\ }\textbf {\bibinfo {volume} {123}},\ \bibinfo
  {pages} {022402} (\bibinfo {year} {2023})}\BibitemShut {NoStop}%
\bibitem [{\citenamefont {Ma}\ and\ \citenamefont {Jia}(2024)}]{Ma2024:PRB}%
  \BibitemOpen
  \bibfield  {author} {\bibinfo {author} {\bibfnamefont {H.-Y.}\ \bibnamefont
  {Ma}}\ and\ \bibinfo {author} {\bibfnamefont {J.-F.}\ \bibnamefont {Jia}},\
  }\bibfield  {title} {\bibinfo {title} {Altermagnetic topological insulator
  and the selection rules},\ }\href
  {https://doi.org/10.1103/PhysRevB.110.064426} {\bibfield  {journal} {\bibinfo
   {journal} {Phys. Rev. B}\ }\textbf {\bibinfo {volume} {110}},\ \bibinfo
  {pages} {064426} (\bibinfo {year} {2024})}\BibitemShut {NoStop}%
\bibitem [{\citenamefont {Tan}\ \emph {et~al.}(2025)\citenamefont {Tan},
  \citenamefont {Gao}, \citenamefont {Yang}, \citenamefont {Liu}, \citenamefont
  {Liu}, \citenamefont {Guo},\ and\ \citenamefont {Lu}}]{Tan2025:PRB}%
  \BibitemOpen
  \bibfield  {author} {\bibinfo {author} {\bibfnamefont {C.-Y.}\ \bibnamefont
  {Tan}}, \bibinfo {author} {\bibfnamefont {Z.-F.}\ \bibnamefont {Gao}},
  \bibinfo {author} {\bibfnamefont {H.-C.}\ \bibnamefont {Yang}}, \bibinfo
  {author} {\bibfnamefont {Z.-X.}\ \bibnamefont {Liu}}, \bibinfo {author}
  {\bibfnamefont {K.}~\bibnamefont {Liu}}, \bibinfo {author} {\bibfnamefont
  {P.-J.}\ \bibnamefont {Guo}},\ and\ \bibinfo {author} {\bibfnamefont {Z.-Y.}\
  \bibnamefont {Lu}},\ }\bibfield  {title} {\bibinfo {title} {Crystal valley
  {{Hall}} effect},\ }\href {https://doi.org/10.1103/PhysRevB.111.094411}
  {\bibfield  {journal} {\bibinfo  {journal} {Phys. Rev. B}\ }\textbf {\bibinfo
  {volume} {111}},\ \bibinfo {pages} {094411} (\bibinfo {year}
  {2025})}\BibitemShut {NoStop}%
\bibitem [{\citenamefont {Feng}\ \emph {et~al.}(2025)\citenamefont {Feng},
  \citenamefont {Tan}, \citenamefont {Gao}, \citenamefont {Yan}, \citenamefont
  {Liu}, \citenamefont {Guo}, \citenamefont {Ma},\ and\ \citenamefont
  {Lu}}]{Feng2025:arXiv}%
  \BibitemOpen
  \bibfield  {author} {\bibinfo {author} {\bibfnamefont {P.}~\bibnamefont
  {Feng}}, \bibinfo {author} {\bibfnamefont {C.-Y.}\ \bibnamefont {Tan}},
  \bibinfo {author} {\bibfnamefont {M.}~\bibnamefont {Gao}}, \bibinfo {author}
  {\bibfnamefont {X.-W.}\ \bibnamefont {Yan}}, \bibinfo {author} {\bibfnamefont
  {Z.-X.}\ \bibnamefont {Liu}}, \bibinfo {author} {\bibfnamefont {P.-J.}\
  \bibnamefont {Guo}}, \bibinfo {author} {\bibfnamefont {F.}~\bibnamefont
  {Ma}},\ and\ \bibinfo {author} {\bibfnamefont {Z.-Y.}\ \bibnamefont {Lu}},\
  }\bibfield  {title} {\bibinfo {title} {Type-{{II}} quantum spin {{Hall}}
  insulator},\ }\href {10.48550/arXiv.2503.13397} {\bibfield  {journal}
  {\bibinfo  {journal} {arXiv:2503.13397}\ } (\bibinfo {year}
  {2025})}\BibitemShut {NoStop}%
\bibitem [{\citenamefont {Antonenko}\ \emph {et~al.}(2025)\citenamefont
  {Antonenko}, \citenamefont {Fernandes},\ and\ \citenamefont
  {Venderbos}}]{Antonenko2025:PRL}%
  \BibitemOpen
  \bibfield  {author} {\bibinfo {author} {\bibfnamefont {D.~S.}\ \bibnamefont
  {Antonenko}}, \bibinfo {author} {\bibfnamefont {R.~M.}\ \bibnamefont
  {Fernandes}},\ and\ \bibinfo {author} {\bibfnamefont {J.~W.~F.}\ \bibnamefont
  {Venderbos}},\ }\bibfield  {title} {\bibinfo {title} {Mirror chern bands and
  weyl nodal loops in altermagnets},\ }\href
  {https://doi.org/10.1103/PhysRevLett.134.096703} {\bibfield  {journal}
  {\bibinfo  {journal} {Phys. Rev. Lett.}\ }\textbf {\bibinfo {volume} {134}},\
  \bibinfo {pages} {096703} (\bibinfo {year} {2025})}\BibitemShut {NoStop}%
\bibitem [{\citenamefont {Zhang}\ \emph
  {et~al.}(2025{\natexlab{b}})\citenamefont {Zhang}, \citenamefont {Cui},
  \citenamefont {Wang}, \citenamefont {Duan}, \citenamefont {Yu},\ and\
  \citenamefont {Yao}}]{Zhang2025:arXiv}%
  \BibitemOpen
  \bibfield  {author} {\bibinfo {author} {\bibfnamefont {R.-W.}\ \bibnamefont
  {Zhang}}, \bibinfo {author} {\bibfnamefont {C.}~\bibnamefont {Cui}}, \bibinfo
  {author} {\bibfnamefont {Y.}~\bibnamefont {Wang}}, \bibinfo {author}
  {\bibfnamefont {J.}~\bibnamefont {Duan}}, \bibinfo {author} {\bibfnamefont
  {Z.-M.}\ \bibnamefont {Yu}},\ and\ \bibinfo {author} {\bibfnamefont
  {Y.}~\bibnamefont {Yao}},\ }\bibfield  {title} {\bibinfo {title} {Quantized
  spin-hall conductivity in altermagnet {Fe$_2$Te$_2$O} with mirror-spin
  coupling},\ }\href {10.48550/arXiv.2503.10681} {\bibfield  {journal}
  {\bibinfo  {journal} {arXiv.2503.10681}\ } (\bibinfo {year}
  {2025}{\natexlab{b}})}\BibitemShut {NoStop}%
\bibitem [{\citenamefont {Shi}\ \emph {et~al.}(2026)\citenamefont {Shi},
  \citenamefont {Jiang}, \citenamefont {Tian}, \citenamefont {Wang},
  \citenamefont {Li}, \citenamefont {Gong},\ and\ \citenamefont
  {Kong}}]{Shi2026:APL}%
  \BibitemOpen
  \bibfield  {author} {\bibinfo {author} {\bibfnamefont {H.}~\bibnamefont
  {Shi}}, \bibinfo {author} {\bibfnamefont {Y.}~\bibnamefont {Jiang}}, \bibinfo
  {author} {\bibfnamefont {Y.}~\bibnamefont {Tian}}, \bibinfo {author}
  {\bibfnamefont {W.}~\bibnamefont {Wang}}, \bibinfo {author} {\bibfnamefont
  {S.}~\bibnamefont {Li}}, \bibinfo {author} {\bibfnamefont {W.-J.}\
  \bibnamefont {Gong}},\ and\ \bibinfo {author} {\bibfnamefont
  {X.}~\bibnamefont {Kong}},\ }\bibfield  {title} {\bibinfo {title} {Tunable
  quantum layer spin {{Hall}} effect in bilayer altermagnetic {Nb$_2$SeTeO}},\
  }\href {https://doi.org/10.1063/5.0312073} {\bibfield  {journal} {\bibinfo
  {journal} {Appl. Phys. Lett.}\ }\textbf {\bibinfo {volume} {128}},\ \bibinfo
  {pages} {063101} (\bibinfo {year} {2026})}\BibitemShut {NoStop}%
\bibitem [{\citenamefont {Fu}\ \emph {et~al.}(2025{\natexlab{b}})\citenamefont
  {Fu}, \citenamefont {Hu}, \citenamefont {Li}, \citenamefont {Duan},
  \citenamefont {Liu},\ and\ \citenamefont {Ouyang}}]{Fu2025:arXiv}%
  \BibitemOpen
  \bibfield  {author} {\bibinfo {author} {\bibfnamefont {Z.}~\bibnamefont
  {Fu}}, \bibinfo {author} {\bibfnamefont {M.}~\bibnamefont {Hu}}, \bibinfo
  {author} {\bibfnamefont {A.}~\bibnamefont {Li}}, \bibinfo {author}
  {\bibfnamefont {H.}~\bibnamefont {Duan}}, \bibinfo {author} {\bibfnamefont
  {J.}~\bibnamefont {Liu}},\ and\ \bibinfo {author} {\bibfnamefont
  {F.}~\bibnamefont {Ouyang}},\ }\bibfield  {title} {\bibinfo {title} {Multiple
  {{Topological Phases Controlled}} via {{Strain}} in {{Two-Dimensional
  Altermagnets}}},\ }\href {10.48550/arXiv.2507.22474} {\bibfield  {journal}
  {\bibinfo  {journal} {arXiv.2507.22474}\ } (\bibinfo {year}
  {2025}{\natexlab{b}})}\BibitemShut {NoStop}%
\bibitem [{\citenamefont {Chen}\ \emph
  {et~al.}(2025{\natexlab{a}})\citenamefont {Chen}, \citenamefont {Zhan},
  \citenamefont {Qin}, \citenamefont {Ma}, \citenamefont {Xu},\ and\
  \citenamefont {Wang}}]{Chen2025:arXiv}%
  \BibitemOpen
  \bibfield  {author} {\bibinfo {author} {\bibfnamefont {Z.}~\bibnamefont
  {Chen}}, \bibinfo {author} {\bibfnamefont {F.}~\bibnamefont {Zhan}}, \bibinfo
  {author} {\bibfnamefont {Z.}~\bibnamefont {Qin}}, \bibinfo {author}
  {\bibfnamefont {D.-S.}\ \bibnamefont {Ma}}, \bibinfo {author} {\bibfnamefont
  {D.-H.}\ \bibnamefont {Xu}},\ and\ \bibinfo {author} {\bibfnamefont
  {R.}~\bibnamefont {Wang}},\ }\bibfield  {title} {\bibinfo {title} {Quantum
  spin hall effect with extended topologically protected features in
  altermangetic multilayers},\ }\href {10.48550/arXiv.2508.03580} {\bibfield
  {journal} {\bibinfo  {journal} {arXiv:2508.03580}\ } (\bibinfo {year}
  {2025}{\natexlab{a}})}\BibitemShut {NoStop}%
\bibitem [{\citenamefont {Yan}\ \emph {et~al.}(2026)\citenamefont {Yan},
  \citenamefont {Ling}, \citenamefont {Han}, \citenamefont {Qi},\ and\
  \citenamefont {He}}]{Yan2026:PRB}%
  \BibitemOpen
  \bibfield  {author} {\bibinfo {author} {\bibfnamefont {X.-H.}\ \bibnamefont
  {Yan}}, \bibinfo {author} {\bibfnamefont {Y.-X.}\ \bibnamefont {Ling}},
  \bibinfo {author} {\bibfnamefont {Y.}~\bibnamefont {Han}}, \bibinfo {author}
  {\bibfnamefont {L.}~\bibnamefont {Qi}},\ and\ \bibinfo {author}
  {\bibfnamefont {A.-L.}\ \bibnamefont {He}},\ }\bibfield  {title} {\bibinfo
  {title} {Altermagnetic topological insulators in an anisotropic system},\
  }\href {https://doi.org/10.1103/6tc8-pr5j} {\bibfield  {journal} {\bibinfo
  {journal} {Phys. Rev. B}\ }\textbf {\bibinfo {volume} {113}},\ \bibinfo
  {pages} {035102} (\bibinfo {year} {2026})}\BibitemShut {NoStop}%
\bibitem [{\citenamefont {Yang}\ \emph {et~al.}(2025)\citenamefont {Yang},
  \citenamefont {Huang},\ and\ \citenamefont {Zhang}}]{Yang2025:NL}%
  \BibitemOpen
  \bibfield  {author} {\bibinfo {author} {\bibfnamefont {N.-J.}\ \bibnamefont
  {Yang}}, \bibinfo {author} {\bibfnamefont {Z.}~\bibnamefont {Huang}},\ and\
  \bibinfo {author} {\bibfnamefont {J.-M.}\ \bibnamefont {Zhang}},\ }\bibfield
  {title} {\bibinfo {title} {Spin-selective second-order topological insulators
  enabling cornertronics in two-dimensional altermagnets},\ }\href
  {https://doi.org/10.1021/acs.nanolett.5c03191} {\bibfield  {journal}
  {\bibinfo  {journal} {Nano Lett.}\ }\textbf {\bibinfo {volume} {25}},\
  \bibinfo {pages} {15495} (\bibinfo {year} {2025})}\BibitemShut {NoStop}%
\bibitem [{\citenamefont {Xu}\ \emph {et~al.}(2025)\citenamefont {Xu},
  \citenamefont {Gao},\ and\ \citenamefont {Liu}}]{Xu2025:NSR}%
  \BibitemOpen
  \bibfield  {author} {\bibinfo {author} {\bibfnamefont {R.}~\bibnamefont
  {Xu}}, \bibinfo {author} {\bibfnamefont {Y.}~\bibnamefont {Gao}},\ and\
  \bibinfo {author} {\bibfnamefont {J.}~\bibnamefont {Liu}},\ }\bibfield
  {title} {\bibinfo {title} {Chemical design of monolayer altermagnets},\
  }\href {https://doi.org/10.1093/nsr/nwaf528} {\bibfield  {journal} {\bibinfo
  {journal} {Nati. Sci. Rev.}\ ,\ \bibinfo {pages} {nwaf528}} (\bibinfo {year}
  {2025})}\BibitemShut {NoStop}%
\bibitem [{\citenamefont {Wan}\ \emph {et~al.}(2025)\citenamefont {Wan},
  \citenamefont {Miao}, \citenamefont {Liu},\ and\ \citenamefont
  {Sun}}]{Wan2025:PRB}%
  \BibitemOpen
  \bibfield  {author} {\bibinfo {author} {\bibfnamefont {Y.-H.}\ \bibnamefont
  {Wan}}, \bibinfo {author} {\bibfnamefont {C.-M.}\ \bibnamefont {Miao}},
  \bibinfo {author} {\bibfnamefont {P.-Y.}\ \bibnamefont {Liu}},\ and\ \bibinfo
  {author} {\bibfnamefont {Q.-F.}\ \bibnamefont {Sun}},\ }\bibfield  {title}
  {\bibinfo {title} {{Helical Fermi arc in altermagnetic Weyl semimetal}},\
  }\href {https://doi.org/10.1103/bdwz-kmyb} {\bibfield  {journal} {\bibinfo
  {journal} {Phys. Rev. B}\ }\textbf {\bibinfo {volume} {112}},\ \bibinfo
  {pages} {235411} (\bibinfo {year} {2025})}\BibitemShut {NoStop}%
\bibitem [{\citenamefont {Seyler}\ \emph {et~al.}(2026)\citenamefont {Seyler},
  \citenamefont {Soavi}, \citenamefont {Weber}, \citenamefont {Das},
  \citenamefont {Agarwal}, \citenamefont {Paradisanos}, \citenamefont {Glazov},
  \citenamefont {Dogadov}, \citenamefont {Gucci}, \citenamefont {Cerullo},
  \citenamefont {Conte}, \citenamefont {Biswas}, \citenamefont {Wilhelm},
  \citenamefont {Žutić}, \citenamefont {Denisov}, \citenamefont {Zhou},
  \citenamefont {Zheng}, \citenamefont {Yao}, \citenamefont {Yu}, \citenamefont
  {Cao}, \citenamefont {Waters}, \citenamefont {Yankowitz}, \citenamefont
  {Burkard}, \citenamefont {Denisov}, \citenamefont {Ihn}, \citenamefont
  {Ensslin}, \citenamefont {Gaudreau}, \citenamefont {Boddison-Chouinard},
  \citenamefont {Fedorova}, \citenamefont {Staude}, \citenamefont {Goh},
  \citenamefont {Zhou},\ and\ \citenamefont {Li}}]{Seyler2026:arXiv}%
  \BibitemOpen
  \bibfield  {author} {\bibinfo {author} {\bibfnamefont {K.~L.}\ \bibnamefont
  {Seyler}}, \bibinfo {author} {\bibfnamefont {G.}~\bibnamefont {Soavi}},
  \bibinfo {author} {\bibfnamefont {B.}~\bibnamefont {Weber}}, \bibinfo
  {author} {\bibfnamefont {S.}~\bibnamefont {Das}}, \bibinfo {author}
  {\bibfnamefont {A.}~\bibnamefont {Agarwal}}, \bibinfo {author} {\bibfnamefont
  {I.}~\bibnamefont {Paradisanos}}, \bibinfo {author} {\bibfnamefont {M.~M.}\
  \bibnamefont {Glazov}}, \bibinfo {author} {\bibfnamefont {O.}~\bibnamefont
  {Dogadov}}, \bibinfo {author} {\bibfnamefont {F.}~\bibnamefont {Gucci}},
  \bibinfo {author} {\bibfnamefont {G.}~\bibnamefont {Cerullo}}, \bibinfo
  {author} {\bibfnamefont {S.~D.}\ \bibnamefont {Conte}}, \bibinfo {author}
  {\bibfnamefont {S.}~\bibnamefont {Biswas}}, \bibinfo {author} {\bibfnamefont
  {J.}~\bibnamefont {Wilhelm}}, \bibinfo {author} {\bibfnamefont
  {I.}~\bibnamefont {Žutić}}, \bibinfo {author} {\bibfnamefont {K.~S.}\
  \bibnamefont {Denisov}}, \bibinfo {author} {\bibfnamefont {T.}~\bibnamefont
  {Zhou}}, \bibinfo {author} {\bibfnamefont {H.}~\bibnamefont {Zheng}},
  \bibinfo {author} {\bibfnamefont {W.}~\bibnamefont {Yao}}, \bibinfo {author}
  {\bibfnamefont {H.}~\bibnamefont {Yu}}, \bibinfo {author} {\bibfnamefont
  {T.}~\bibnamefont {Cao}}, \bibinfo {author} {\bibfnamefont {D.}~\bibnamefont
  {Waters}}, \bibinfo {author} {\bibfnamefont {M.}~\bibnamefont {Yankowitz}},
  \bibinfo {author} {\bibfnamefont {G.}~\bibnamefont {Burkard}}, \bibinfo
  {author} {\bibfnamefont {A.}~\bibnamefont {Denisov}}, \bibinfo {author}
  {\bibfnamefont {T.}~\bibnamefont {Ihn}}, \bibinfo {author} {\bibfnamefont
  {K.}~\bibnamefont {Ensslin}}, \bibinfo {author} {\bibfnamefont
  {L.}~\bibnamefont {Gaudreau}}, \bibinfo {author} {\bibfnamefont
  {J.}~\bibnamefont {Boddison-Chouinard}}, \bibinfo {author} {\bibfnamefont
  {Z.}~\bibnamefont {Fedorova}}, \bibinfo {author} {\bibfnamefont
  {I.}~\bibnamefont {Staude}}, \bibinfo {author} {\bibfnamefont {K.~E.~J.}\
  \bibnamefont {Goh}}, \bibinfo {author} {\bibfnamefont {Z.}~\bibnamefont
  {Zhou}},\ and\ \bibinfo {author} {\bibfnamefont {X.}~\bibnamefont {Li}},\
  }\bibfield  {title} {\bibinfo {title} {Valleytronics in {2D} materials
  roadmap},\ }\href {https://arxiv.org/abs/2603.01427} {\bibfield  {journal}
  {\bibinfo  {journal} {arXiv.2603.01427}\ } (\bibinfo {year}
  {2026})}\BibitemShut {NoStop}%
\bibitem [{\citenamefont {Schaibley}\ \emph {et~al.}(2016)\citenamefont
  {Schaibley}, \citenamefont {Yu}, \citenamefont {Clark}, \citenamefont
  {Rivera}, \citenamefont {Ross}, \citenamefont {Seyler}, \citenamefont {Yao},\
  and\ \citenamefont {Xu}}]{schaibley2016:NRM}%
  \BibitemOpen
  \bibfield  {author} {\bibinfo {author} {\bibfnamefont {J.~R.}\ \bibnamefont
  {Schaibley}}, \bibinfo {author} {\bibfnamefont {H.}~\bibnamefont {Yu}},
  \bibinfo {author} {\bibfnamefont {G.}~\bibnamefont {Clark}}, \bibinfo
  {author} {\bibfnamefont {P.}~\bibnamefont {Rivera}}, \bibinfo {author}
  {\bibfnamefont {J.~S.}\ \bibnamefont {Ross}}, \bibinfo {author}
  {\bibfnamefont {K.~L.}\ \bibnamefont {Seyler}}, \bibinfo {author}
  {\bibfnamefont {W.}~\bibnamefont {Yao}},\ and\ \bibinfo {author}
  {\bibfnamefont {X.}~\bibnamefont {Xu}},\ }\bibfield  {title} {\bibinfo
  {title} {Valleytronics in {{2D}} materials},\ }\href
  {https://doi.org/10.1038/natrevmats.2016.55} {\bibfield  {journal} {\bibinfo
  {journal} {Nat. Rev. Mater.}\ }\textbf {\bibinfo {volume} {1}},\ \bibinfo
  {pages} {16055} (\bibinfo {year} {2016})}\BibitemShut {NoStop}%
\bibitem [{\citenamefont {Li}\ \emph {et~al.}(2016)\citenamefont {Li},
  \citenamefont {Wang}, \citenamefont {McFaul}, \citenamefont {Zern},
  \citenamefont {Ren}, \citenamefont {Watanabe}, \citenamefont {Taniguchi},
  \citenamefont {Qiao},\ and\ \citenamefont {Zhu}}]{Li2016:NN}%
  \BibitemOpen
  \bibfield  {author} {\bibinfo {author} {\bibfnamefont {J.}~\bibnamefont
  {Li}}, \bibinfo {author} {\bibfnamefont {K.}~\bibnamefont {Wang}}, \bibinfo
  {author} {\bibfnamefont {K.~J.}\ \bibnamefont {McFaul}}, \bibinfo {author}
  {\bibfnamefont {Z.}~\bibnamefont {Zern}}, \bibinfo {author} {\bibfnamefont
  {Y.}~\bibnamefont {Ren}}, \bibinfo {author} {\bibfnamefont {K.}~\bibnamefont
  {Watanabe}}, \bibinfo {author} {\bibfnamefont {T.}~\bibnamefont {Taniguchi}},
  \bibinfo {author} {\bibfnamefont {Z.}~\bibnamefont {Qiao}},\ and\ \bibinfo
  {author} {\bibfnamefont {J.}~\bibnamefont {Zhu}},\ }\bibfield  {title}
  {\bibinfo {title} {Gate-controlled topological conducting channels in bilayer
  graphene},\ }\href {https://doi.org/10.1038/nnano.2016.158} {\bibfield
  {journal} {\bibinfo  {journal} {Nature Nanotech}\ }\textbf {\bibinfo {volume}
  {11}},\ \bibinfo {pages} {1060} (\bibinfo {year} {2016})}\BibitemShut
  {NoStop}%
\bibitem [{\citenamefont {Li}\ \emph {et~al.}(2018)\citenamefont {Li},
  \citenamefont {Zhang}, \citenamefont {Yin}, \citenamefont {Zhang},
  \citenamefont {Watanabe}, \citenamefont {Taniguchi}, \citenamefont {Liu},\
  and\ \citenamefont {Zhu}}]{Li2018:Science}%
  \BibitemOpen
  \bibfield  {author} {\bibinfo {author} {\bibfnamefont {J.}~\bibnamefont
  {Li}}, \bibinfo {author} {\bibfnamefont {R.-X.}\ \bibnamefont {Zhang}},
  \bibinfo {author} {\bibfnamefont {Z.}~\bibnamefont {Yin}}, \bibinfo {author}
  {\bibfnamefont {J.}~\bibnamefont {Zhang}}, \bibinfo {author} {\bibfnamefont
  {K.}~\bibnamefont {Watanabe}}, \bibinfo {author} {\bibfnamefont
  {T.}~\bibnamefont {Taniguchi}}, \bibinfo {author} {\bibfnamefont
  {C.}~\bibnamefont {Liu}},\ and\ \bibinfo {author} {\bibfnamefont
  {J.}~\bibnamefont {Zhu}},\ }\bibfield  {title} {\bibinfo {title} {A valley
  valve and electron beam splitter},\ }\href
  {https://doi.org/10.1126/science.aao5989} {\bibfield  {journal} {\bibinfo
  {journal} {Science}\ }\textbf {\bibinfo {volume} {362}},\ \bibinfo {pages}
  {1149} (\bibinfo {year} {2018})}\BibitemShut {NoStop}%
\bibitem [{\citenamefont {Ren}\ \emph {et~al.}(2016)\citenamefont {Ren},
  \citenamefont {Qiao},\ and\ \citenamefont {Niu}}]{Ren2016:RPP}%
  \BibitemOpen
  \bibfield  {author} {\bibinfo {author} {\bibfnamefont {Y.}~\bibnamefont
  {Ren}}, \bibinfo {author} {\bibfnamefont {Z.}~\bibnamefont {Qiao}},\ and\
  \bibinfo {author} {\bibfnamefont {Q.}~\bibnamefont {Niu}},\ }\bibfield
  {title} {\bibinfo {title} {Topological phases in two-dimensional materials:
  {{A}} review},\ }\href {https://doi.org/10.1088/0034-4885/79/6/066501}
  {\bibfield  {journal} {\bibinfo  {journal} {Rep. Prog. Phys.}\ }\textbf
  {\bibinfo {volume} {79}},\ \bibinfo {pages} {066501} (\bibinfo {year}
  {2016})}\BibitemShut {NoStop}%
\bibitem [{SM()}]{SM}%
  \BibitemOpen
  \href@noop {} {}\bibinfo {note} {See Supplemental Material for the expanded
  discussion of the tight-binding model, topological characterization and phase
  diagram, disorder-dependent transport simulations, first-principles
  calculations, and material realizations.}\BibitemShut {Stop}%
\bibitem [{\citenamefont {Zhu}\ \emph {et~al.}(2025{\natexlab{b}})\citenamefont
  {Zhu}, \citenamefont {Liu}, \citenamefont {Duan}, \citenamefont {Zhang},
  \citenamefont {Hao}, \citenamefont {Wei}, \citenamefont {{\v{Z}}uti{\'c}},\
  and\ \citenamefont {Zhou}}]{Zhu2025:SCPMA}%
  \BibitemOpen
  \bibfield  {author} {\bibinfo {author} {\bibfnamefont {Z.}~\bibnamefont
  {Zhu}}, \bibinfo {author} {\bibfnamefont {Y.}~\bibnamefont {Liu}}, \bibinfo
  {author} {\bibfnamefont {X.}~\bibnamefont {Duan}}, \bibinfo {author}
  {\bibfnamefont {J.}~\bibnamefont {Zhang}}, \bibinfo {author} {\bibfnamefont
  {B.}~\bibnamefont {Hao}}, \bibinfo {author} {\bibfnamefont {S.-H.}\
  \bibnamefont {Wei}}, \bibinfo {author} {\bibfnamefont {I.}~\bibnamefont
  {{\v{Z}}uti{\'c}}},\ and\ \bibinfo {author} {\bibfnamefont {T.}~\bibnamefont
  {Zhou}},\ }\bibfield  {title} {\bibinfo {title} {Emergent multiferroic
  altermagnets and spin control via noncollinear molecular polarization},\
  }\href {https://doi.org/10.1007/s11433-025-2778-3} {\bibfield  {journal}
  {\bibinfo  {journal} {Sci. China Phys. Mech. Astron.}\ }\textbf {\bibinfo
  {volume} {68}},\ \bibinfo {pages} {127562} (\bibinfo {year}
  {2025}{\natexlab{b}})}\BibitemShut {NoStop}%
\bibitem [{\citenamefont {Zhu}\ \emph {et~al.}(2025{\natexlab{c}})\citenamefont
  {Zhu}, \citenamefont {Chen}, \citenamefont {Duan}, \citenamefont {Cui},
  \citenamefont {Zhang}, \citenamefont {{\v{Z}}uti{\'c}},\ and\ \citenamefont
  {Zhou}}]{Zhu2025:arXiv2}%
  \BibitemOpen
  \bibfield  {author} {\bibinfo {author} {\bibfnamefont {Z.}~\bibnamefont
  {Zhu}}, \bibinfo {author} {\bibfnamefont {X.}~\bibnamefont {Chen}}, \bibinfo
  {author} {\bibfnamefont {X.}~\bibnamefont {Duan}}, \bibinfo {author}
  {\bibfnamefont {Z.}~\bibnamefont {Cui}}, \bibinfo {author} {\bibfnamefont
  {J.}~\bibnamefont {Zhang}}, \bibinfo {author} {\bibfnamefont
  {I.}~\bibnamefont {{\v{Z}}uti{\'c}}},\ and\ \bibinfo {author} {\bibfnamefont
  {T.}~\bibnamefont {Zhou}},\ }\bibfield  {title} {\bibinfo {title}
  {Altermagnetoelectric spin field effect transistor},\ }\href
  {https://arxiv.org/abs/2512.02974} {\bibfield  {journal} {\bibinfo  {journal}
  {arXiv:2512.02974}\ } (\bibinfo {year} {2025}{\natexlab{c}})}\BibitemShut
  {NoStop}%
\bibitem [{\citenamefont {Liu}\ \emph {et~al.}(2025{\natexlab{b}})\citenamefont
  {Liu}, \citenamefont {Shao}, \citenamefont {Li}, \citenamefont {Wan},
  \citenamefont {Chen},\ and\ \citenamefont {Xing}}]{Liu2025:arxiv}%
  \BibitemOpen
  \bibfield  {author} {\bibinfo {author} {\bibfnamefont {L.-S.}\ \bibnamefont
  {Liu}}, \bibinfo {author} {\bibfnamefont {K.}~\bibnamefont {Shao}}, \bibinfo
  {author} {\bibfnamefont {H.-D.}\ \bibnamefont {Li}}, \bibinfo {author}
  {\bibfnamefont {X.}~\bibnamefont {Wan}}, \bibinfo {author} {\bibfnamefont
  {W.}~\bibnamefont {Chen}},\ and\ \bibinfo {author} {\bibfnamefont
  {D.}~\bibnamefont {Xing}},\ }\bibfield  {title} {\bibinfo {title}
  {Altermagnetic spin precession and spin transistor},\ }\href
  {https://arxiv.org/abs/2511.05208} {\bibfield  {journal} {\bibinfo  {journal}
  {arXiv:2511.05208}\ } (\bibinfo {year} {2025}{\natexlab{b}})}\BibitemShut
  {NoStop}%
\bibitem [{\citenamefont {Zhang}\ \emph {et~al.}(2013)\citenamefont {Zhang},
  \citenamefont {MacDonald},\ and\ \citenamefont {Mele}}]{Zhang2013:PNAS}%
  \BibitemOpen
  \bibfield  {author} {\bibinfo {author} {\bibfnamefont {F.}~\bibnamefont
  {Zhang}}, \bibinfo {author} {\bibfnamefont {A.~H.}\ \bibnamefont
  {MacDonald}},\ and\ \bibinfo {author} {\bibfnamefont {E.~J.}\ \bibnamefont
  {Mele}},\ }\bibfield  {title} {\bibinfo {title} {Valley {{Chern}} numbers and
  boundary modes in gapped bilayer graphene},\ }\href
  {https://doi.org/10.1073/pnas.1308853110} {\bibfield  {journal} {\bibinfo
  {journal} {Proc. Natl. Acad. Sci.}\ }\textbf {\bibinfo {volume} {110}},\
  \bibinfo {pages} {10546} (\bibinfo {year} {2013})}\BibitemShut {NoStop}%
\bibitem [{\citenamefont {Sheng}\ \emph {et~al.}(2006)\citenamefont {Sheng},
  \citenamefont {Weng}, \citenamefont {Sheng},\ and\ \citenamefont
  {Haldane}}]{Sheng2006:PRL}%
  \BibitemOpen
  \bibfield  {author} {\bibinfo {author} {\bibfnamefont {D.~N.}\ \bibnamefont
  {Sheng}}, \bibinfo {author} {\bibfnamefont {Z.~Y.}\ \bibnamefont {Weng}},
  \bibinfo {author} {\bibfnamefont {L.}~\bibnamefont {Sheng}},\ and\ \bibinfo
  {author} {\bibfnamefont {F.~D.~M.}\ \bibnamefont {Haldane}},\ }\bibfield
  {title} {\bibinfo {title} {Quantum {{Spin-Hall Effect}} and {{Topologically
  Invariant Chern Numbers}}},\ }\href
  {https://doi.org/10.1103/PhysRevLett.97.036808} {\bibfield  {journal}
  {\bibinfo  {journal} {Phys. Rev. Lett.}\ }\textbf {\bibinfo {volume} {97}},\
  \bibinfo {pages} {036808} (\bibinfo {year} {2006})}\BibitemShut {NoStop}%
\bibitem [{\citenamefont {Zhou}\ \emph {et~al.}(2015)\citenamefont {Zhou},
  \citenamefont {Zhang}, \citenamefont {Zhao}, \citenamefont {Zhang},\ and\
  \citenamefont {Yang}}]{Zhou2015:NL}%
  \BibitemOpen
  \bibfield  {author} {\bibinfo {author} {\bibfnamefont {T.}~\bibnamefont
  {Zhou}}, \bibinfo {author} {\bibfnamefont {J.}~\bibnamefont {Zhang}},
  \bibinfo {author} {\bibfnamefont {B.}~\bibnamefont {Zhao}}, \bibinfo {author}
  {\bibfnamefont {H.}~\bibnamefont {Zhang}},\ and\ \bibinfo {author}
  {\bibfnamefont {Z.}~\bibnamefont {Yang}},\ }\bibfield  {title} {\bibinfo
  {title} {Quantum spin-quantum anomalous {{Hall}} insulators and topological
  transitions in functionalized {{Sb}}(111) monolayers},\ }\href
  {https://doi.org/10.1021/acs.nanolett.5b01373} {\bibfield  {journal}
  {\bibinfo  {journal} {Nano Lett.}\ }\textbf {\bibinfo {volume} {15}},\
  \bibinfo {pages} {5149} (\bibinfo {year} {2015})}\BibitemShut {NoStop}%
\bibitem [{\citenamefont {Zhou}\ \emph {et~al.}(2016)\citenamefont {Zhou},
  \citenamefont {Zhang}, \citenamefont {Xue}, \citenamefont {Zhao},
  \citenamefont {Zhang}, \citenamefont {Jiang},\ and\ \citenamefont
  {Yang}}]{Zhou2016:PRB}%
  \BibitemOpen
  \bibfield  {author} {\bibinfo {author} {\bibfnamefont {T.}~\bibnamefont
  {Zhou}}, \bibinfo {author} {\bibfnamefont {J.}~\bibnamefont {Zhang}},
  \bibinfo {author} {\bibfnamefont {Y.}~\bibnamefont {Xue}}, \bibinfo {author}
  {\bibfnamefont {B.}~\bibnamefont {Zhao}}, \bibinfo {author} {\bibfnamefont
  {H.}~\bibnamefont {Zhang}}, \bibinfo {author} {\bibfnamefont
  {H.}~\bibnamefont {Jiang}},\ and\ \bibinfo {author} {\bibfnamefont
  {Z.}~\bibnamefont {Yang}},\ }\bibfield  {title} {\bibinfo {title} {{Quantum
  spin-quantum anomalous Hall effect with tunable edge states in {Sb}
  monolayer-based heterostructures}},\ }\href
  {https://doi.org/10.1103/PhysRevB.94.235449} {\bibfield  {journal} {\bibinfo
  {journal} {Phys. Rev. B}\ }\textbf {\bibinfo {volume} {94}},\ \bibinfo
  {pages} {235449} (\bibinfo {year} {2016})}\BibitemShut {NoStop}%
\bibitem [{\citenamefont {Gonz\'alez-Hern\'andez}\ \emph
  {et~al.}(2025)\citenamefont {Gonz\'alez-Hern\'andez}, \citenamefont
  {Serrano},\ and\ \citenamefont {Uribe}}]{Rafael2025:PRB}%
  \BibitemOpen
  \bibfield  {author} {\bibinfo {author} {\bibfnamefont {R.}~\bibnamefont
  {Gonz\'alez-Hern\'andez}}, \bibinfo {author} {\bibfnamefont {H.}~\bibnamefont
  {Serrano}},\ and\ \bibinfo {author} {\bibfnamefont {B.}~\bibnamefont
  {Uribe}},\ }\bibfield  {title} {\bibinfo {title} {Spin chern number in
  altermagnets},\ }\href {https://doi.org/10.1103/PhysRevB.111.085127}
  {\bibfield  {journal} {\bibinfo  {journal} {Phys. Rev. B}\ }\textbf {\bibinfo
  {volume} {111}},\ \bibinfo {pages} {085127} (\bibinfo {year}
  {2025})}\BibitemShut {NoStop}%
\bibitem [{\citenamefont {Sancho}\ \emph {et~al.}(1984)\citenamefont {Sancho},
  \citenamefont {Sancho},\ and\ \citenamefont {Rubio}}]{Sancho1984:JPFMP}%
  \BibitemOpen
  \bibfield  {author} {\bibinfo {author} {\bibfnamefont {M.~P.~L.}\
  \bibnamefont {Sancho}}, \bibinfo {author} {\bibfnamefont {J.~M.~L.}\
  \bibnamefont {Sancho}},\ and\ \bibinfo {author} {\bibfnamefont
  {J.}~\bibnamefont {Rubio}},\ }\bibfield  {title} {\bibinfo {title} {Quick
  iterative scheme for the calculation of transfer matrices: Application to
  {{Mo}} (100)},\ }\href {https://doi.org/10.1088/0305-4608/14/5/016}
  {\bibfield  {journal} {\bibinfo  {journal} {J. Phys. F: Met. Phys.}\ }\textbf
  {\bibinfo {volume} {14}},\ \bibinfo {pages} {1205} (\bibinfo {year}
  {1984})}\BibitemShut {NoStop}%
\bibitem [{\citenamefont {Jiang}\ \emph {et~al.}(2009)\citenamefont {Jiang},
  \citenamefont {Wang}, \citenamefont {Sun},\ and\ \citenamefont
  {Xie}}]{Jiang2009:PRB}%
  \BibitemOpen
  \bibfield  {author} {\bibinfo {author} {\bibfnamefont {H.}~\bibnamefont
  {Jiang}}, \bibinfo {author} {\bibfnamefont {L.}~\bibnamefont {Wang}},
  \bibinfo {author} {\bibfnamefont {Q.-F.}\ \bibnamefont {Sun}},\ and\ \bibinfo
  {author} {\bibfnamefont {X.~C.}\ \bibnamefont {Xie}},\ }\bibfield  {title}
  {\bibinfo {title} {Numerical study of the topological {{Anderson}} insulator
  in {{HgTe}}/{{CdTe}} quantum wells},\ }\href
  {https://doi.org/10.1103/PhysRevB.80.165316} {\bibfield  {journal} {\bibinfo
  {journal} {Phys. Rev. B}\ }\textbf {\bibinfo {volume} {80}},\ \bibinfo
  {pages} {165316} (\bibinfo {year} {2009})}\BibitemShut {NoStop}%
\bibitem [{\citenamefont {K{\"o}nig}\ \emph {et~al.}(2007)\citenamefont
  {K{\"o}nig}, \citenamefont {Wiedmann}, \citenamefont {Br{\"u}ne},
  \citenamefont {Roth}, \citenamefont {Buhmann}, \citenamefont {Molenkamp},
  \citenamefont {Qi},\ and\ \citenamefont {Zhang}}]{Konig2007:Science}%
  \BibitemOpen
  \bibfield  {author} {\bibinfo {author} {\bibfnamefont {M.}~\bibnamefont
  {K{\"o}nig}}, \bibinfo {author} {\bibfnamefont {S.}~\bibnamefont {Wiedmann}},
  \bibinfo {author} {\bibfnamefont {C.}~\bibnamefont {Br{\"u}ne}}, \bibinfo
  {author} {\bibfnamefont {A.}~\bibnamefont {Roth}}, \bibinfo {author}
  {\bibfnamefont {H.}~\bibnamefont {Buhmann}}, \bibinfo {author} {\bibfnamefont
  {L.~W.}\ \bibnamefont {Molenkamp}}, \bibinfo {author} {\bibfnamefont {X.-L.}\
  \bibnamefont {Qi}},\ and\ \bibinfo {author} {\bibfnamefont {S.-C.}\
  \bibnamefont {Zhang}},\ }\bibfield  {title} {\bibinfo {title} {Quantum {{Spin
  Hall Insulator State}} in {{HgTe Quantum Wells}}},\ }\href
  {https://doi.org/10.1126/science.1148047} {\bibfield  {journal} {\bibinfo
  {journal} {Science}\ }\textbf {\bibinfo {volume} {318}},\ \bibinfo {pages}
  {766} (\bibinfo {year} {2007})}\BibitemShut {NoStop}%
\bibitem [{\citenamefont {Ozawa}\ \emph {et~al.}(2019)\citenamefont {Ozawa},
  \citenamefont {Price}, \citenamefont {Amo}, \citenamefont {Goldman},
  \citenamefont {Hafezi}, \citenamefont {Lu}, \citenamefont {Rechtsman},
  \citenamefont {Schuster}, \citenamefont {Simon}, \citenamefont {Zilberberg},\
  and\ \citenamefont {Carusotto}}]{Ozawa2019:RMP}%
  \BibitemOpen
  \bibfield  {author} {\bibinfo {author} {\bibfnamefont {T.}~\bibnamefont
  {Ozawa}}, \bibinfo {author} {\bibfnamefont {H.~M.}\ \bibnamefont {Price}},
  \bibinfo {author} {\bibfnamefont {A.}~\bibnamefont {Amo}}, \bibinfo {author}
  {\bibfnamefont {N.}~\bibnamefont {Goldman}}, \bibinfo {author} {\bibfnamefont
  {M.}~\bibnamefont {Hafezi}}, \bibinfo {author} {\bibfnamefont
  {L.}~\bibnamefont {Lu}}, \bibinfo {author} {\bibfnamefont {M.~C.}\
  \bibnamefont {Rechtsman}}, \bibinfo {author} {\bibfnamefont {D.}~\bibnamefont
  {Schuster}}, \bibinfo {author} {\bibfnamefont {J.}~\bibnamefont {Simon}},
  \bibinfo {author} {\bibfnamefont {O.}~\bibnamefont {Zilberberg}},\ and\
  \bibinfo {author} {\bibfnamefont {I.}~\bibnamefont {Carusotto}},\ }\bibfield
  {title} {\bibinfo {title} {Topological photonics},\ }\href
  {https://doi.org/10.1103/RevModPhys.91.015006} {\bibfield  {journal}
  {\bibinfo  {journal} {Rev. Mod. Phys.}\ }\textbf {\bibinfo {volume} {91}},\
  \bibinfo {pages} {015006} (\bibinfo {year} {2019})}\BibitemShut {NoStop}%
\bibitem [{\citenamefont {Zhu}\ \emph {et~al.}(2023)\citenamefont {Zhu},
  \citenamefont {Deng}, \citenamefont {Liu}, \citenamefont {Lu}, \citenamefont
  {Wang}, \citenamefont {Lin}, \citenamefont {Huang}, \citenamefont {Jiang},\
  and\ \citenamefont {Liu}}]{zhu2023:RPP}%
  \BibitemOpen
  \bibfield  {author} {\bibinfo {author} {\bibfnamefont {W.}~\bibnamefont
  {Zhu}}, \bibinfo {author} {\bibfnamefont {W.}~\bibnamefont {Deng}}, \bibinfo
  {author} {\bibfnamefont {Y.}~\bibnamefont {Liu}}, \bibinfo {author}
  {\bibfnamefont {J.}~\bibnamefont {Lu}}, \bibinfo {author} {\bibfnamefont
  {H.-X.}\ \bibnamefont {Wang}}, \bibinfo {author} {\bibfnamefont {Z.-K.}\
  \bibnamefont {Lin}}, \bibinfo {author} {\bibfnamefont {X.}~\bibnamefont
  {Huang}}, \bibinfo {author} {\bibfnamefont {J.-H.}\ \bibnamefont {Jiang}},\
  and\ \bibinfo {author} {\bibfnamefont {Z.}~\bibnamefont {Liu}},\ }\bibfield
  {title} {\bibinfo {title} {Topological phononic metamaterials},\ }\href
  {https://doi.org/10.1088/1361-6633/aceeee} {\bibfield  {journal} {\bibinfo
  {journal} {Rep. Prog. Phys.}\ }\textbf {\bibinfo {volume} {86}},\ \bibinfo
  {pages} {106501} (\bibinfo {year} {2023})}\BibitemShut {NoStop}%
\bibitem [{\citenamefont {Chen}\ \emph
  {et~al.}(2025{\natexlab{b}})\citenamefont {Chen}, \citenamefont {Liu},
  \citenamefont {Liu}, \citenamefont {Yu}, \citenamefont {Ren}, \citenamefont
  {Li}, \citenamefont {Zhang},\ and\ \citenamefont {Liu}}]{Chen2025:Nature}%
  \BibitemOpen
  \bibfield  {author} {\bibinfo {author} {\bibfnamefont {X.}~\bibnamefont
  {Chen}}, \bibinfo {author} {\bibfnamefont {Y.}~\bibnamefont {Liu}}, \bibinfo
  {author} {\bibfnamefont {P.}~\bibnamefont {Liu}}, \bibinfo {author}
  {\bibfnamefont {Y.}~\bibnamefont {Yu}}, \bibinfo {author} {\bibfnamefont
  {J.}~\bibnamefont {Ren}}, \bibinfo {author} {\bibfnamefont {J.}~\bibnamefont
  {Li}}, \bibinfo {author} {\bibfnamefont {A.}~\bibnamefont {Zhang}},\ and\
  \bibinfo {author} {\bibfnamefont {Q.}~\bibnamefont {Liu}},\ }\bibfield
  {title} {\bibinfo {title} {Unconventional magnons in collinear magnets
  dictated by spin space groups},\ }\href
  {https://doi.org/10.1038/s41586-025-08715-7} {\bibfield  {journal} {\bibinfo
  {journal} {Nature}\ }\textbf {\bibinfo {volume} {640}},\ \bibinfo {pages}
  {349} (\bibinfo {year} {2025}{\natexlab{b}})}\BibitemShut {NoStop}%
\bibitem [{\citenamefont {Zhu}\ \emph {et~al.}(2025{\natexlab{d}})\citenamefont
  {Zhu}, \citenamefont {Huang}, \citenamefont {Chen}, \citenamefont {Duan},
  \citenamefont {Zhang}, \citenamefont {{\v{Z}}uti{\'c}},\ and\ \citenamefont
  {Zhou}}]{Zhu2025:arXiv}%
  \BibitemOpen
  \bibfield  {author} {\bibinfo {author} {\bibfnamefont {Z.}~\bibnamefont
  {Zhu}}, \bibinfo {author} {\bibfnamefont {R.}~\bibnamefont {Huang}}, \bibinfo
  {author} {\bibfnamefont {X.}~\bibnamefont {Chen}}, \bibinfo {author}
  {\bibfnamefont {X.}~\bibnamefont {Duan}}, \bibinfo {author} {\bibfnamefont
  {J.}~\bibnamefont {Zhang}}, \bibinfo {author} {\bibfnamefont
  {I.}~\bibnamefont {{\v{Z}}uti{\'c}}},\ and\ \bibinfo {author} {\bibfnamefont
  {T.}~\bibnamefont {Zhou}},\ }\bibfield  {title} {\bibinfo {title}
  {Altermagnetic {{Proximity Effect}}},\ }\href
  {https://arxiv.org/abs/2509.06790} {\bibfield  {journal} {\bibinfo  {journal}
  {arXiv.2509.06790}\ } (\bibinfo {year} {2025}{\natexlab{d}})}\BibitemShut
  {NoStop}%
\bibitem [{\citenamefont {Zhou}\ \emph {et~al.}(2020)\citenamefont {Zhou},
  \citenamefont {Dartiailh}, \citenamefont {Mayer}, \citenamefont {Han},
  \citenamefont {Matos-Abiague}, \citenamefont {Shabani},\ and\ \citenamefont
  {\ifmmode \check{Z}\else \v{Z}\fi{}uti\ifmmode~\acute{c}\else
  \'{c}\fi{}}}]{Zhou2020:PRL}%
  \BibitemOpen
  \bibfield  {author} {\bibinfo {author} {\bibfnamefont {T.}~\bibnamefont
  {Zhou}}, \bibinfo {author} {\bibfnamefont {M.~C.}\ \bibnamefont {Dartiailh}},
  \bibinfo {author} {\bibfnamefont {W.}~\bibnamefont {Mayer}}, \bibinfo
  {author} {\bibfnamefont {J.~E.}\ \bibnamefont {Han}}, \bibinfo {author}
  {\bibfnamefont {A.}~\bibnamefont {Matos-Abiague}}, \bibinfo {author}
  {\bibfnamefont {J.}~\bibnamefont {Shabani}},\ and\ \bibinfo {author}
  {\bibfnamefont {I.}~\bibnamefont {\ifmmode \check{Z}\else
  \v{Z}\fi{}uti\ifmmode~\acute{c}\else \'{c}\fi{}}},\ }\bibfield  {title}
  {\bibinfo {title} {Phase control of {Majorana} bound states in a topological
  $\mathsf{X}$ junction},\ }\href
  {https://doi.org/10.1103/PhysRevLett.124.137001} {\bibfield  {journal}
  {\bibinfo  {journal} {Phys. Rev. Lett.}\ }\textbf {\bibinfo {volume} {124}},\
  \bibinfo {pages} {137001} (\bibinfo {year} {2020})}\BibitemShut {NoStop}%
\bibitem [{\citenamefont {Amundsen}\ \emph {et~al.}(2024)\citenamefont
  {Amundsen}, \citenamefont {Linder}, \citenamefont {Robinson}, \citenamefont
  {\ifmmode \check{Z}\else \v{Z}\fi{}uti\ifmmode~\acute{c}\else \'{c}\fi{}},\
  and\ \citenamefont {Banerjee}}]{Amundsen2024:RMP}%
  \BibitemOpen
  \bibfield  {author} {\bibinfo {author} {\bibfnamefont {M.}~\bibnamefont
  {Amundsen}}, \bibinfo {author} {\bibfnamefont {J.}~\bibnamefont {Linder}},
  \bibinfo {author} {\bibfnamefont {J.~W.~A.}\ \bibnamefont {Robinson}},
  \bibinfo {author} {\bibfnamefont {I.}~\bibnamefont {\ifmmode \check{Z}\else
  \v{Z}\fi{}uti\ifmmode~\acute{c}\else \'{c}\fi{}}},\ and\ \bibinfo {author}
  {\bibfnamefont {N.}~\bibnamefont {Banerjee}},\ }\bibfield  {title} {\bibinfo
  {title} {Colloquium: {{Spin-orbit}} effects in superconducting hybrid
  structures},\ }\href {https://doi.org/10.1103/RevModPhys.96.021003}
  {\bibfield  {journal} {\bibinfo  {journal} {Rev. Mod. Phys.}\ }\textbf
  {\bibinfo {volume} {96}},\ \bibinfo {pages} {021003} (\bibinfo {year}
  {2024})}\BibitemShut {NoStop}%
\bibitem [{\citenamefont {{\v{Z}}uti{\'c}}\ \emph {et~al.}(2004)\citenamefont
  {{\v{Z}}uti{\'c}}, \citenamefont {Fabian},\ and\ \citenamefont {{Das
  Sarma}}}]{vZutic2004:RMP}%
  \BibitemOpen
  \bibfield  {author} {\bibinfo {author} {\bibfnamefont {I.}~\bibnamefont
  {{\v{Z}}uti{\'c}}}, \bibinfo {author} {\bibfnamefont {J.}~\bibnamefont
  {Fabian}},\ and\ \bibinfo {author} {\bibfnamefont {S.}~\bibnamefont {{Das
  Sarma}}},\ }\bibfield  {title} {\bibinfo {title} {Spintronics:
  {{Fundamentals}} and applications},\ }\href
  {https://link.aps.org/doi/10.1103/RevModPhys.76.323} {\bibfield  {journal}
  {\bibinfo  {journal} {Rev. Mod. Phys.}\ }\textbf {\bibinfo {volume} {76}},\
  \bibinfo {pages} {323} (\bibinfo {year} {2004})}\BibitemShut {NoStop}%
\end{thebibliography}%

\end{document}